\documentclass{spi-b2111}
\def\today{31 November 2015}
\usepackage{ws-rv-van}   
 \def\ArtDir{} 
\def \D {\hbox{d}} 
\begin{document}
\chapter{Exact solution of the planar motion of~three\\ ~arbitrary point vortices} 
{\vglue -5.0truemm} 

\author{R.~Conte and L.~de Seze} 

\address{Service de physique du solide et de r\'esonance magn\'etique, \\ CEN Saclay, France. DPhG/PSRM/1697/80} 

\begin{abstract}
We give an exact quantitative solution for the motion of three vortices of any strength,
which Poincar\'{e} showed to be integrable. The absolute motion of one vortex is
generally biperiodic: in uniformly rotating axes, the motion is periodic. There are two
kinds of relative equilibrium configuration: two equilateral triangles and one or three
colinear configurations, their stability conditions split the strengths space into three
domains in which the sets of trajectories are topologically distinct.

According to the values of the strengths and the initial positions, all possible 
motions are classified. Two sets of strengths lead to generic motions other than
biperiodic. First, when the angular momentum vanishes, besides the biperiodic regime
there exists an expansion spiral motion and even a triple collision in a finite time, but
the latter motion is nongeneric. Second, when two strengths are opposite, the system also
exhibits the elastic diffusion of a vortex doublet by the third vortex.

For given values of the invariants, the volume of the phase space of this
Hamiltonian system is proportional to the period of the reduced motion, a well
known result of the theory of adiabatic invariants. We then formally examine the
behaviour of the quantities that Onsager defined only for a large number of
interacting vortices. 
\end{abstract}
\section*{Introduction}

Known from a long time, the problem of the motion of a system of point vortices in
interaction presents a particular interest for the study of bidimensional
turbulence. As predicted by Onsager (1949) the study of the thermodynamics of a
large system of vortices shows the possibility of
\hbox{negative}\vadjust{\vspace*{25pt}\pagebreak} temperatures and eddy
viscosities (see e.g. Lundgren and Pointin 1977). The case of a small number of
vortices was studied, long ago, by Lord Kelvin and Mayer (1878). They used
experimental methods to obtain some results on the stability of simple geometric
configurations. Recently Novikov (1975) pointed out the interest of the three
vortex system as the first elementary interaction process in isotropic turbulence
kinetics. Using properties of the triangle he was able to solve the problem of
three identical vortices.

While this work was under submission, the referees pointed out to us the existence
of a paper to appear. In this paper, Hassan Aref (March 1979), extending the work
of Novikov with a symmetry-preserving presentation, qualitatively solved the
relative motion of three vortices and undertook a classification of the topology
of the phase space; however, he gave no indication on the nature of the absolute
motion and no quantitative results, except for two special cases of great
importance which he solved completely: the direct or exchange scattering when two
strengths are equal and opposite to the third one, and the self-similar motion of
a triple collision or a triple expansion to infinity when both the inertia
momentum and the angular momentum vanish.

In this paper, we present a method which quantitatively gives the absolute motion
in all cases of strengths or initial conditions. The main idea is to introduce
reduced variables with a very simple geometrical interpretation and whose number
matches the number of degrees of freedom of the system. For three vortices, this
can be done by defining one complex variable $\zeta$ which characterizes the shape
of the triangle. Knowing the motion of $\zeta$, which happens to be periodic, is
then sufficient to derive the behaviour of any physically interesting variable. We
thus have obtained detailed results, which resort mainly to fluid mechanics and
partly to the field of differentiable dynamical systems.

In the first part, we describe the problem, introduce the reduced motion and
discuss its advantages and inconvenients. In the second part, we give the general
solution: generically, i.e. for arbitrary strengths and initial positions, the
absolute motion of any given vortex is the product of a periodic motion of period
$T$ by a uniform rotation or, stated in other words, the motion is periodic when
referred to uniformly rotating axes centered at the center of vorticity; in the
case of a vortex plasma (zero total strength), the uniform rotation is merely
replaced by a uniform translation. Due to its frequent occurrence in nature, we
give special consideration to the case of two or three equal strengths: in a given
domain of initial conditions, vortices with equal strengths have the same motion,
up to a shift of $\frac{T}{2}$ or $\frac{T}{3}$ in time and a rotation in space.

The third part, following the Smale's method of study of a differentiable
dynamical system, is devoted to the finding of the bifurcation set, i.e. the set
of values of the invariants for which the nature of the motion changes. This
reduces to the finding of the relative equilibria, which are shown to be of two
types, the same than those of the three body problem of celestial mechanics: two
equilateral triangles and one or three colinear configurations. The number and
stability of these relative equilibria are discussed, thus leading to a separation
of the strengths space in three principal domains where the sets of orbits are
topologically different; an important result is that, whatever be the strengths
domain, the phase space is always multiconnected; for strengths on the boundaries
of these domains, the motion must be studied separately for it presents special
features.

In the fourth part, we study the absolute motion whenever it differs from the general
biperiodic case, i.e. when strengths are on the boundaries of their limiting domains or
when initial positions are those of a relative equilibrium. At a relative equilibrium the
motion is a rigid body rotation. Otherwise two new generic motions are found. First, for
strengths such that the angular momentum vanishes, the vortices can go to infinity in a
spiral motion with a fixed asymptotic shape of the triangle; there even exists a spiral
motion ending in a triple collision in a finite time, but it is nongeneric since it
happens only for a zero value of the inertia momentum. Second, when two strengths are
opposite, besides the doubly periodic motion an elastic scattering can occur in which a
vortex doublet is diffused by the third vortex; if moreover the third vortex has the
strength of one of the two others, a third generic motion exists which is an exchange
scattering.

The fifth part consists in formally examining the behaviour of the thermodynamical
quantities, of course meaningless for this integrable system, that Onsager defined
for a large number of vortices; the volume of the phase space is found to be equal
to the period of the relative motion; among the features which could be indicative
for a large number of vortices is a possible lack of ergodicity due to a multiple
connexity of the phase space.

\enlargethispage{-12pt}

\section{Description of the problem}

We consider three point vortices $M_{j}$ $(j=1,2,3)$ in a plane with strengths
$\kappa_{j}$ and given initial positions. Their motion is ruled by the first order
\hbox{differential} system:
\begin{equation}
\forall\,j=1,2,3\enskip {:}\enskip 2\pi i\frac{\D\bar{z}_{j}}{\D t}=\sum_{\substack{\ell=1\\
\ell \ne j}}^{3}\frac{\kappa_{\ell}}{z_{j}-z_{\ell}},
\end{equation}
where $z=x\,{+}\,iy$ and the bar denotes the complex conjugation. This system is
equivalent to the set of Hamilton equations for the Hamiltonian:
\begin{equation}
H=-\frac{1}{2\pi}\sum_{\substack{j\\
j}}\vphantom{\sum}_{\substack{\\ \\ \\ \\ <}}\sum_{\substack{\ell\\[2pt]
\ell}}\kappa_{j}\kappa_{\ell}\enskip{\rm Log}|z_{j}-z_{\ell}|,
\end{equation}
in the conjugate variables $\sqrt{|\kappa_{j}|}x_{j}$ and $\sqrt{|\kappa_{j}|}$
sign $(\kappa_{j})y_{j}$, $j=1,2,3$; therefore $H$ is an invariant and remains
equal to the energy $E$. The invariance of $H$ under translation and rotation
yields two other invariants:
\begin{eqnarray*}
&&\hbox{the impulse } B=\sum_{j}\kappa_{j}z_{j}=X+iY,\\[4pt]
&& I=\sum_{j}\kappa_{j}|z_{j}|^{2},
\end{eqnarray*}
which Poincar\'{e} called inertia momentum.

In fact we shall use, instead of $I$, the invariant
\begin{equation}
J=\sum_{\substack{j\\
j}}\vphantom{\sum}_{\substack{\\ \\ \\ \\ <}}\sum_{\substack{\ell\\[2pt]
\ell}}\kappa_{j}\kappa_{\ell}|z_{j}-z_{\ell}|^{2}=(\kappa_{1}+\kappa_{2}+\kappa_{3})I-|B|^{2},
\end{equation}
which depends only on the relative positions of the vortices. There are $6-4=2$ remaining
degrees of freedom and we suppose the problem to be nondegenerate:
$\kappa_{1}\kappa_{2}\kappa_{3}(z_{2}-z_{3})(z_{3}-z_{1})(z_{1}-z_{2})\ne 0$.

Among the six Poisson brackets built from the four known integrals of motion $H,J,X,Y$,
only one is nonzero:
\[
\{X,Y\}=\sum_{j}\kappa_{j}=K.
\]

Since a Hamiltonian system with $2N$ variables is integrable in the sense of Liouville
when it has $N$ independent invariants in involution, the three vortex system $(N=3)$ is
therefore integrable (Poincar\'{e}, 1893). The purpose of this paper is to integrate it.

Let us remark that this problem is not affected by the adjunction of an external
velocity field made of a uniform translation and a uniform rotation, since the new
equations of motion
\[
2\pi
i\frac{\D\bar{z}_{j}}{\D t}=\sum_{\ell}'\frac{\kappa_{\ell}}{z_{j}-z_{\ell}}+2\pi
i\bar{v}+2\pi \omega\bar{z}_{j},\quad \omega \in {\mathbb R},\quad v \in  {\mathbb
C}
\]
reduce to the original ones by the change of variables
\[
Z_{j}=\frac{iv}{\omega}+\left(z_{j}-\frac{iv}{\omega}\right)e^{-i\omega t}.
\]

\subsection*{The reduced motion}

The main idea is to match the number of variables and the number of degrees of
freedom of this system, so as to keep the minimum number of independent variables.
For this purpose, let us define a time dependent complex plane $\zeta$ in which
two of the three point vortices remain fixed. This can be achieved by the
following transformation
\begin{equation}
z\to
\zeta=\frac{(\kappa_{2}+\kappa_{3})z-(\kappa_{2}z_{2}+\kappa_{3}z_{3})}{z_{2}-z_{3}},
\end{equation}
where 2 and 3 number two vortices whose sum of strengths is nonzero; the affixes of these
two vortices become $\kappa_{3}$ and $-\kappa_{2}$ under the transformation.

It is clearly seen from the definition how a reduced point is geometrically
deduced from an absolute point. We shall simply note $\zeta$ the transformed
of~$M_{1}$:
\[
\zeta=\xi+i\eta=\frac{Kz_{1}-B}{z_{2}-z_{3}},
\]
$\zeta$ therefore represents the shape of the triangle, and the inverse
transformation is represented by
 \setcounter{equation}{4}
\begin{gather}
z_{1}=z_{1},\quad z_{2}=\frac{(\kappa_{3}K-\kappa_{1}\zeta)z_{1}+(\zeta-\kappa_{3})B}{s\zeta},\nn\\[4pt]
z_{3}=\frac{(-\kappa_{2}K-\kappa_{1}\zeta)z_{1}+(\zeta+\kappa_{2})B}{s\zeta},
\end{gather}
where $s=\kappa_{2}+\kappa_{3}\ne 0$.

The reader will have noticed the main disadvantage of the above definition of two reduced
coordinates $\xi,\eta$: it does not reflect the invariance of the problem under the
permutations of the three elements $(\kappa_{j},z_{j})$ and therefore every result we can
get may be uneasy to interpret. Nevertheless, the advantages are numerous. First, every
physical quantity can systematically be expressed, as we shall soon see, as a function of
$\zeta$ and $Kz_{1}-B$ only; moreover, since $\zeta$ is invariant under a change of
length and $Kz_{1}-B$ extensive in the lengths, such a physical function will quite
generally be the product of a function of $\zeta$ by a function of $Kz_{1}-B$ and we are
going to see that this uncoupling between intensive and extensive variables will enable
us to solve the motion not only qualitatively but also quantitatively. Secondly, unlike
Novikov and Aref, we do not have to eliminate some unphysical portions of our $\zeta$
plane (which will be seen to be the initial conditions plane) since every $\zeta$ point
describes a physical situation. Thirdly, this reduction leaves only the required number
of degrees of freedom.

\section{The solution for the general case}

To obtain the absolute motion we need only determine the evolution of $\zeta$ and
$z_{1}$, since we have the parametric representation (5). Provided $K$ and $B$ do
not simultaneously vanish the motions of $z_{1}$ and $\zeta$ are ruled by:
\begin{eqnarray}
2\pi i\frac{\D\bar{z}_{1}}{\D t}&=&\frac{1}{Kz_{1}-B}\frac{s^{2}\zeta(\zeta+d)}{(\zeta+\kappa_{2})(\zeta-\kappa_{3})},\\[3pt]
2\pi i\frac{\D\bar{\zeta}}{\D t}&=&\frac{-1}{|Kz_{1}-B|^{2}}
\frac{s|\zeta|^{2}[\bar{\zeta}(\zeta^{2}+d\zeta-Q)-sK(\zeta+d)]}
{(\zeta+\kappa_{2})(\zeta-\kappa_{3})},
\end{eqnarray}
where $d=\kappa_{2}-\kappa_{3}$ and
$Q=\kappa_{1}\kappa_{2}+\kappa_{2}\kappa_{3}+\kappa_{3}\kappa_{1}$.

We can express the invariants $E$ and $J$ as functions of $z_{1}$ and $\zeta$:
\begin{gather}
J=|Kz_{1}-B|^{2}\frac{\kappa_{1}|\zeta|^{2}+\kappa_{2}\kappa_{3}K}{s|\zeta|^{2}},\\
e^{-4\pi
E}=|Kz_{1}-B|^{2Q}|\zeta|^{-2Q}\left|\frac{\zeta-\kappa_{3}}{s}\right|^{2\kappa_{1}\kappa_{2}}
 \left|\frac{\zeta+\kappa_{2}}{s}\right|^{2\kappa_{1}\kappa_{3}},
\end{gather}
expressions where we notice the factorized dependency on $\zeta$ and $z_{1}$.

Unless $Q$ and $J$ simultaneously vanish, at least one of the two above equations
expresses $|Kz_{1}-B|$ as a function of $\zeta$ and from (7) we obtain a first order
differential system for $\zeta$. For example if $J$ is nonzero the elimination of $z_{1}$
between (7) and (8) gives:
\begin{equation}
2\pi
i\frac{\D\bar{\zeta}}{\D t}=\frac{-(\kappa_{1}|\zeta|^{2}+\kappa_{2}\kappa_{3}K)[\bar{\zeta}(\zeta^{2}+d\zeta-Q)-sK(\zeta+d)]}
{J(\zeta+\kappa_{2})(\zeta-\kappa_{3})}.
\end{equation}

\removelastskip\pagebreak

\noindent There is no need to solve this system since the equation of the reduced
trajectory is given by the straightforward elimination of $z_{1}$ between (8)
and~(9):
\begin{equation}
\frac{\kappa_{1}|\zeta|^{2}+\kappa_{2}\kappa_{3}K}{sQ}\left|\frac{\zeta-\kappa_{3}}{s}\right|^{-2\frac{\kappa_{1}\kappa_{2}}{Q}}
\left|\frac{\zeta+\kappa_{2}}{s}\right|^{-2\frac{\kappa_{1}\kappa_{3}}{Q}}=\frac{J}{Q}e^{4\pi\frac{E}{Q}}.
\end{equation}
This represents in the $\zeta$ plane a set of closed orbits, indexed by the
non-dimensional variable $c=\frac{J}{Q}e^{\frac{4\pi E}{Q}}$ which is invariant
under a change of length or a change of unit of vorticity.

\begin{figure}[b]
\centerline{\includegraphics{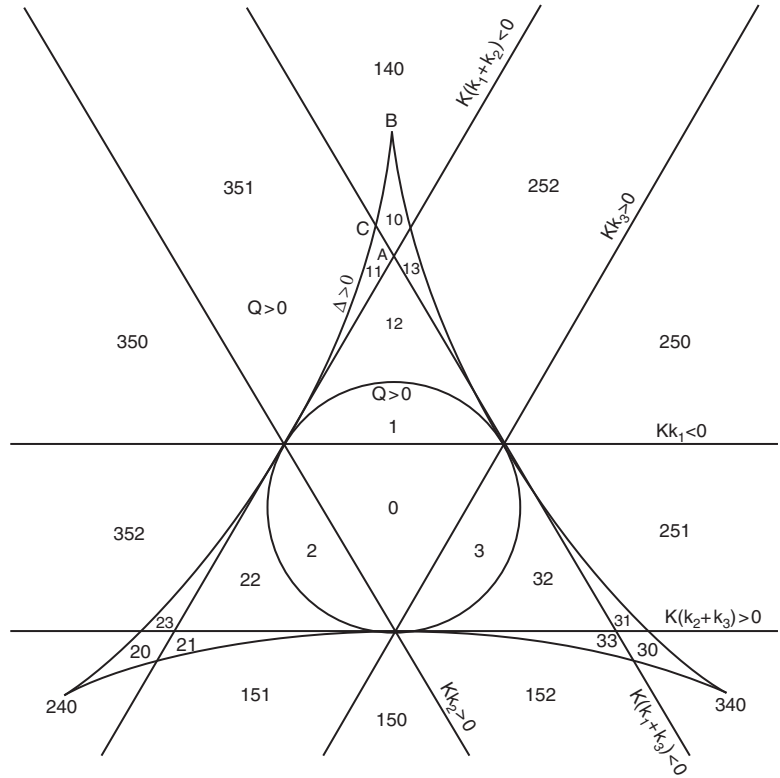}} 

 \caption{The strengths space. The numbering of regions reflects the ternary
symmetry.\label{fig1}}
\end{figure}

\begin{figure}
 \centerline{\includegraphics{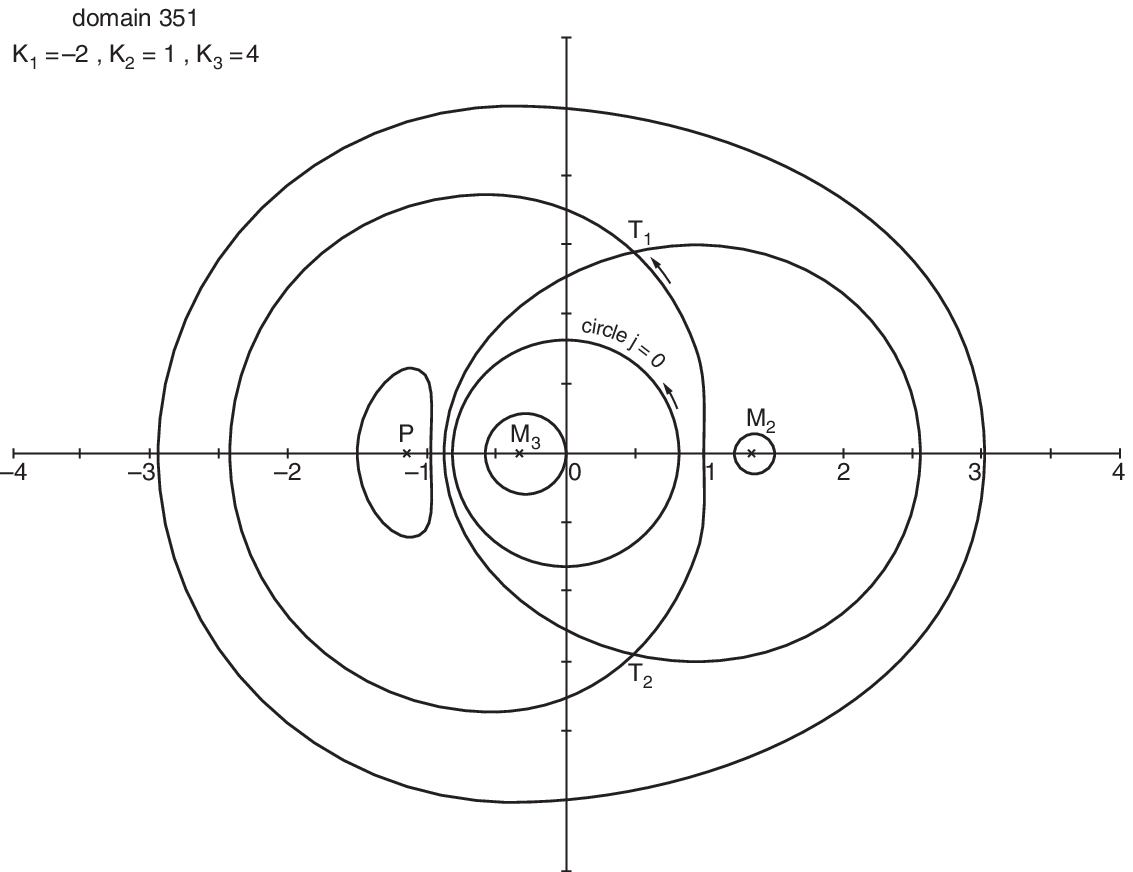}}
 \scaption{$\zeta$ plane in the domain 351 ($Q<0$, $\Delta > 0$)\newline
 domain 351\newline
$\kappa_{1}=-2,\kappa_{2}=1,\kappa_{3}=4$.\label{fig2}}
\end{figure}

\begin{figure}
 \centerline{\includegraphics{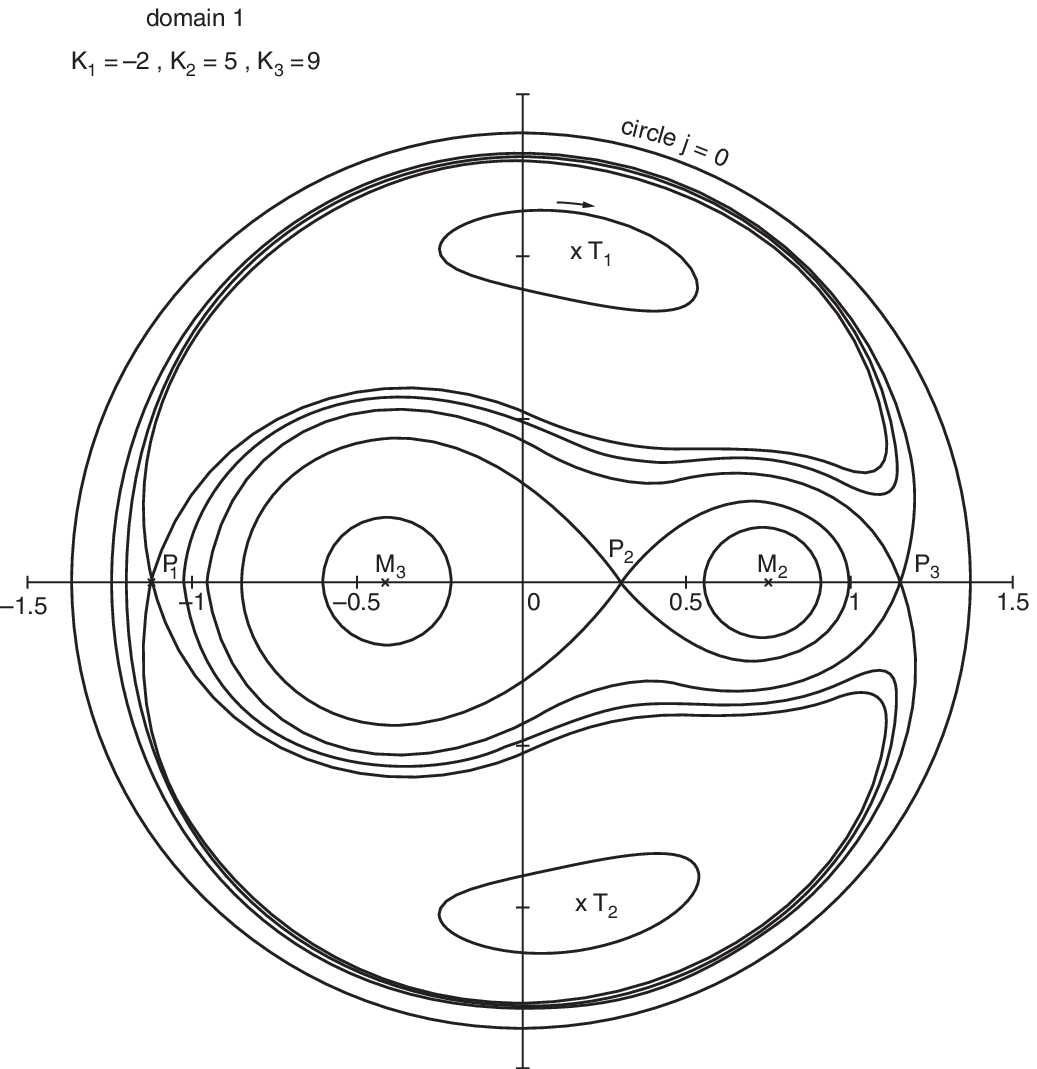}}
 \scaption{$\zeta$ plane in the domain 1 ($Q>0$, $\Delta < 0$) domain 1\newline
$\kappa_{1}=-2,\kappa_{2}=5,\kappa_{3}=9$.\label{fig3}}
\end{figure}

\begin{figure}
 \centerline{\includegraphics{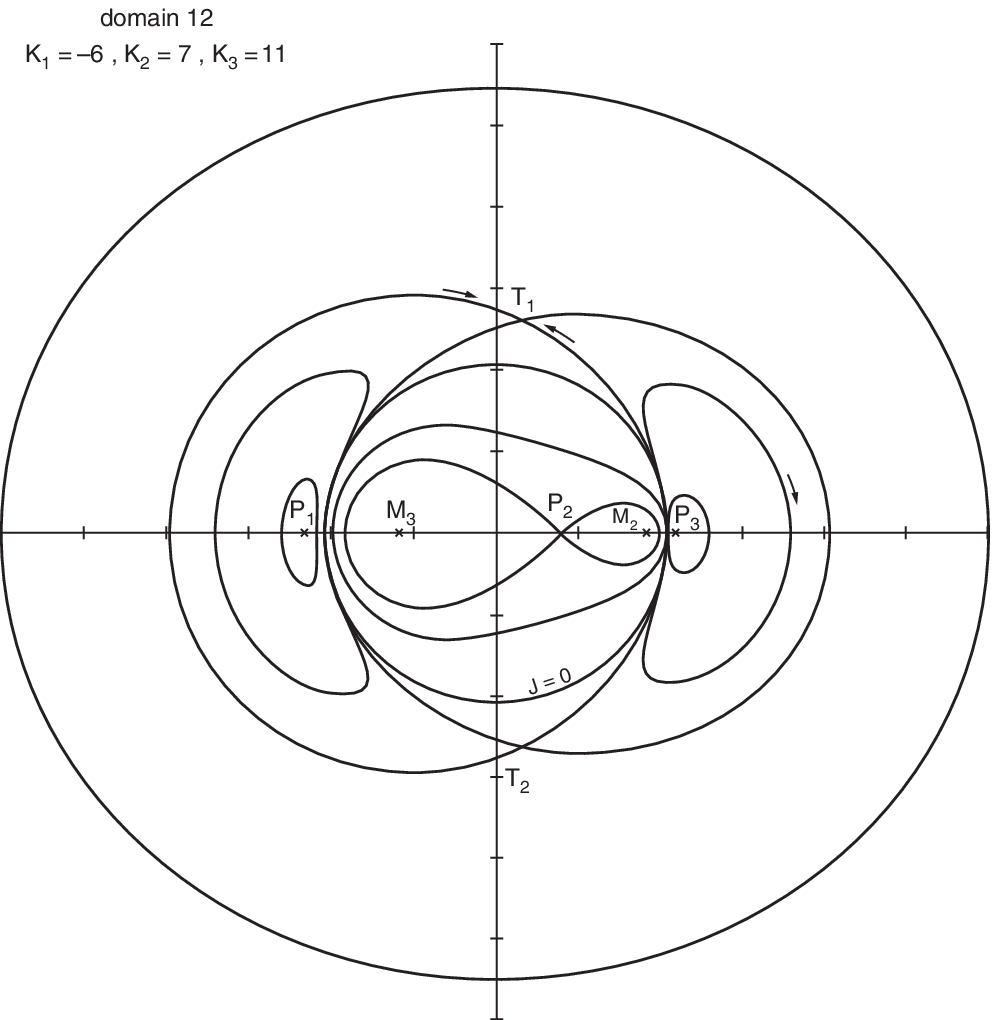}}
 \scaption{$\zeta$ plane in the domain 12 ($Q<0$, $\Delta < 0$) domain 12\newline
$\kappa_{1}=-6,\kappa_{2}=7,\kappa_{3}=11$.\label{fig4}}
\end{figure}

\begin{figure}
 \centerline{\includegraphics{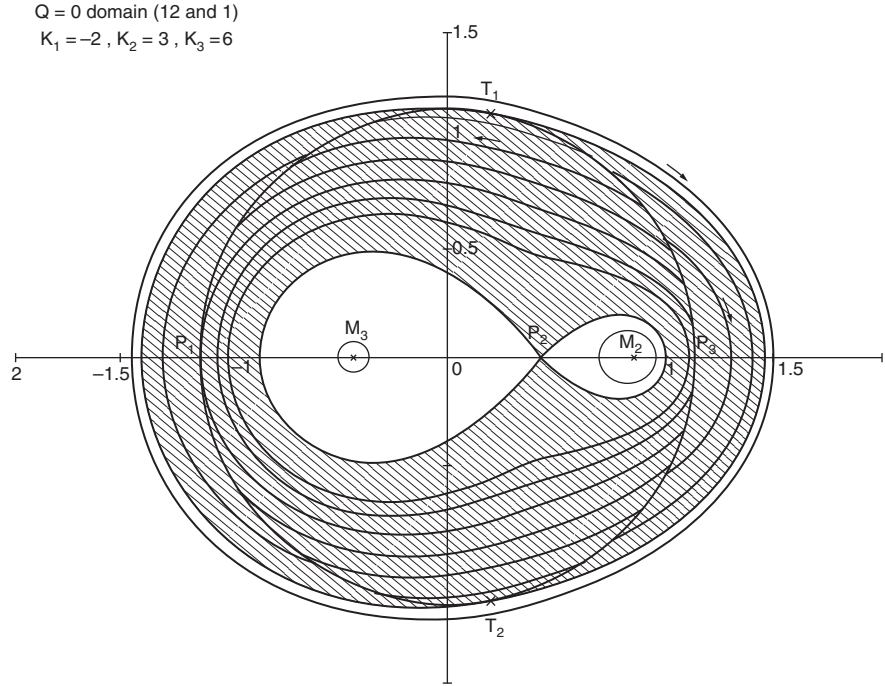}}
\caption{$\zeta$ plane on the line $Q=0$. The nonperiodic domain is hatched $Q=0$ domain
(12 and 1) $\kappa_{1}=-2,\kappa_{2}=3,\kappa_{3}=6$.\label{fig5}}
\end{figure}

\begin{figure}
 \centerline{\includegraphics{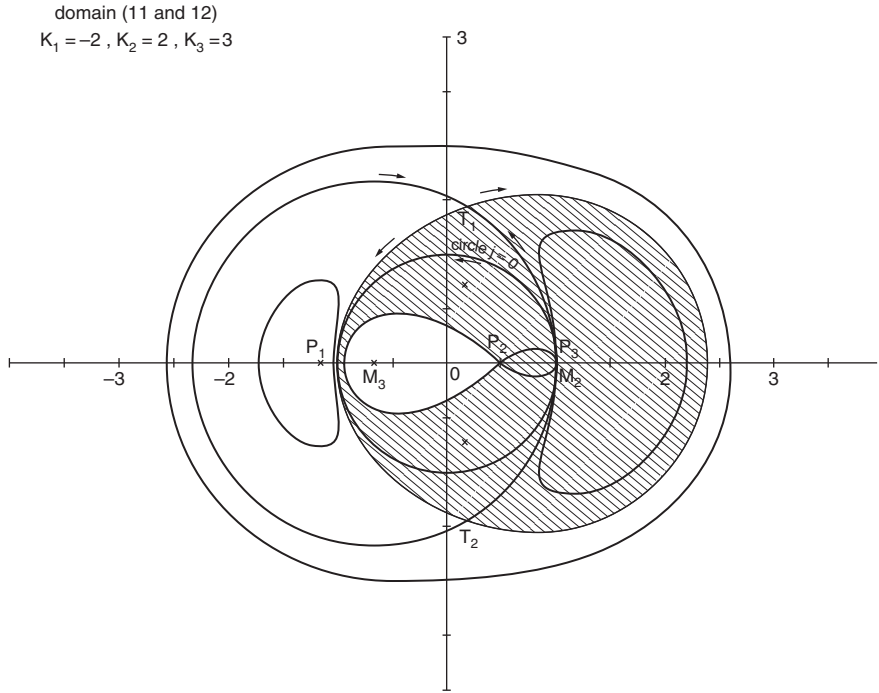}}
 \caption{$\zeta$ plane on the line (11 and 12). The diffusion domain is hatched
domain (11 and 12) $\kappa_{1}=-2,\kappa_{2}=2,\kappa_{3}=3$.\label{fig6}}
 \vspace*{-5pt}
\end{figure}

The set of orbits is symmetrical relative to the $\xi$ axis and also, when
$\kappa_{2}$ equals $\kappa_{3}$, to the $\eta$ axis. When two vortices are close
to each other, they remain as such and therefore the $\zeta$ curves are near to
circles in the vicinity of $-\kappa_{2}, \kappa_{3}$ and $\infty$.
Figures~\ref{fig2} to \ref{fig6} show examples of the $\zeta$ plane.

\setcounter{sch}{1}

Since there is in general no stationary point on the orbit, the reduced motion is
periodic and the period is expressed by
\begin{equation}
T=-\frac{\pi J}{s}\oint\frac{|(\zeta+\kappa_{2})(\zeta-\kappa_{3})|^{2}}
{(\kappa_{1}|\zeta|^{2}+\kappa_{2}\kappa_{3}K)^{2}}\frac{\D |\zeta|^{2}}{{\rm
Im}(\zeta^{2}+d\zeta)}.
\end{equation}
When $J$ is zero we use (9) instead of (8) to obtain a similar expression, a
result which shows that the condition $J=0$ alone represents nothing special as
one would believe in the Aref classification (for more details see the third part
of this paper). To have a dimensionless result we can take as unit of time
$T_{u}=\frac{4\pi^{2}J}{QK}$ which is, as we shall see later, the period of the
absolute motion when the vortices are in a configuration of relative equilibrium;
therefore, in every domain of the multiply-connected $\zeta$ plane,
$\frac{T}{T_{u}}$ depends only on $c$.

\subsection*{The absolute motion}

Using the parametric representation of the $z_{j}$'s we easily obtain the
following relations:
\begin{equation}
\frac{Kz_{1}-B}{s\zeta}=\frac{Kz_{2}-B}{\kappa_{3}K-\kappa_{1}\zeta}=
\frac{Kz_{3}-B}{-\kappa_{2}K-\kappa_{1}\zeta}.
\end{equation}

Then, using (8) or (9), we conclude that, when the reduced motion is periodic, the
modulus of $z_{j}-\frac{B}{K}$ has the period $T$ of the reduced motion. As to the
arguments of these affixes, after one period they have all been increased by the
same value, modulo $2\pi$:
\begin{equation}
\left\{\begin{array}{@{}l@{}}
\ds \Delta\varphi_{1}=\left[\arg\left(z_{1}-\frac{B}{K}\right)\right]_{o}^{T}\\[14pt]
\ds \qquad\;=\oint \frac{-K {\rm Re}\{(\zeta^{2}+d\zeta)(\bar{\zeta}^{2}+d\bar{\zeta}-\kappa_{2}\kappa_{3})\}}{2|\zeta|^{2}(\kappa_{1}|\zeta|^{2}+\kappa_{2}\kappa_{3}K){\rm Im}(\zeta^{2}+d\zeta)}\D|\zeta|^{2}=\Delta\varphi,\\[14pt]
\ds \Delta\varphi_{2}=\Delta\varphi_{1}+\left[\arg\left(\frac{\kappa_{3}K-\kappa_{1}\zeta}{s\zeta}\right)\right]_{o}^{T}=\Delta \varphi\ {\rm modulo}\ 2\pi,\\[14pt]
\ds \Delta\varphi_{3}=\Delta\varphi_{1}+\left[\arg\left(\frac{-\kappa_{2}K-\kappa_{1}\zeta}{s\zeta}\right)\right]_{o}^{T}=\Delta \varphi\ {\rm modulo}\ 2\pi.\\
\end{array}\right.
\end{equation}

For the motion it means that, after a time interval of $T$, the shape and size of the
triangle are again the same, i.e. the new positions are deduced from their initial values
by a rotation of $\Delta \varphi$ around the center of vorticity. Depending on the domain
of initial conditions, the number of turns around the barycentrum may vary from one
vortex to another by an integer value. Therefore, when $K$ is nonzero, the absolute
motion of any vortex is the product of a uniform rotation about the barycentrum and of a
periodic motion, the two periods being the same for the three vortices:
\[
\forall\,j=1,2,3\enskip{:}\enskip
z_{j}(t)=\frac{B}{K}+\left(z_{j}(0)-\frac{B}{K}\right)e^{i\Delta
\varphi\frac{t}{T}}f_{j}\left(\frac{t}{T}\right),
\]
where $f_{j}$ is periodic with period 1. In other words, in uniformly rotating
axes centered at the barycentrum, the absolute motion is periodic.
Figure~\ref{fig8} shows the absolute trajectory of $M_{1}$ both in fixed axes and
in rotating axes, for $\vec{\kappa}=(-2,1,4)$ and $\zeta_{o}=-\frac{3K}{2}$ (the
$\zeta$ orbit is the small curve surrounding $P$ in Figure~\ref{fig2}).

\fig{6}
\begin{figure}
 \centerline{\includegraphics{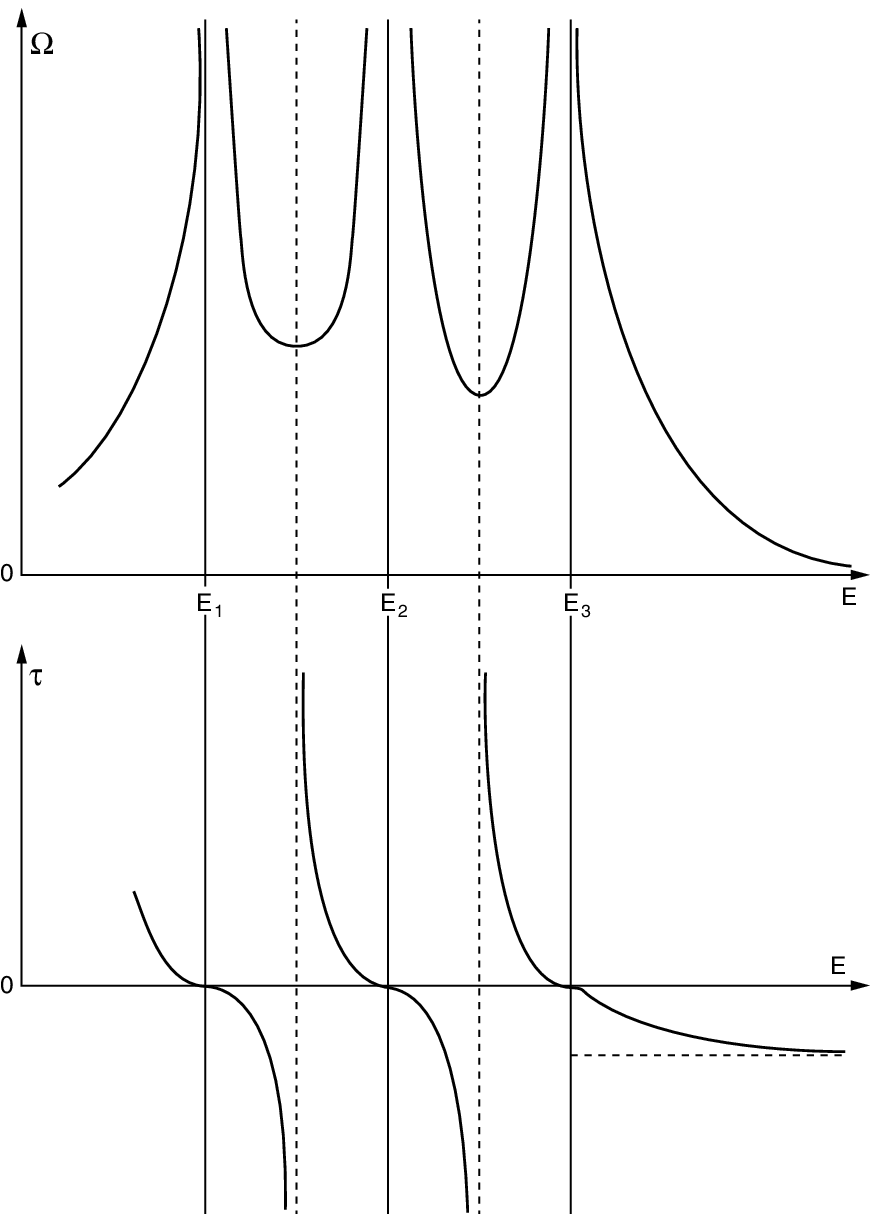}}
 \caption{Volume $\Omega$ and temperature $\tau$ versus energy in the domain $0(k_{j}K>0)$.
Data for these qualitative figures are taken from
$\vec{\kappa}=(2,9,13)$.\label{fig7}}
\end{figure}

\begin{figure}[!t]
 \centerline{\includegraphics{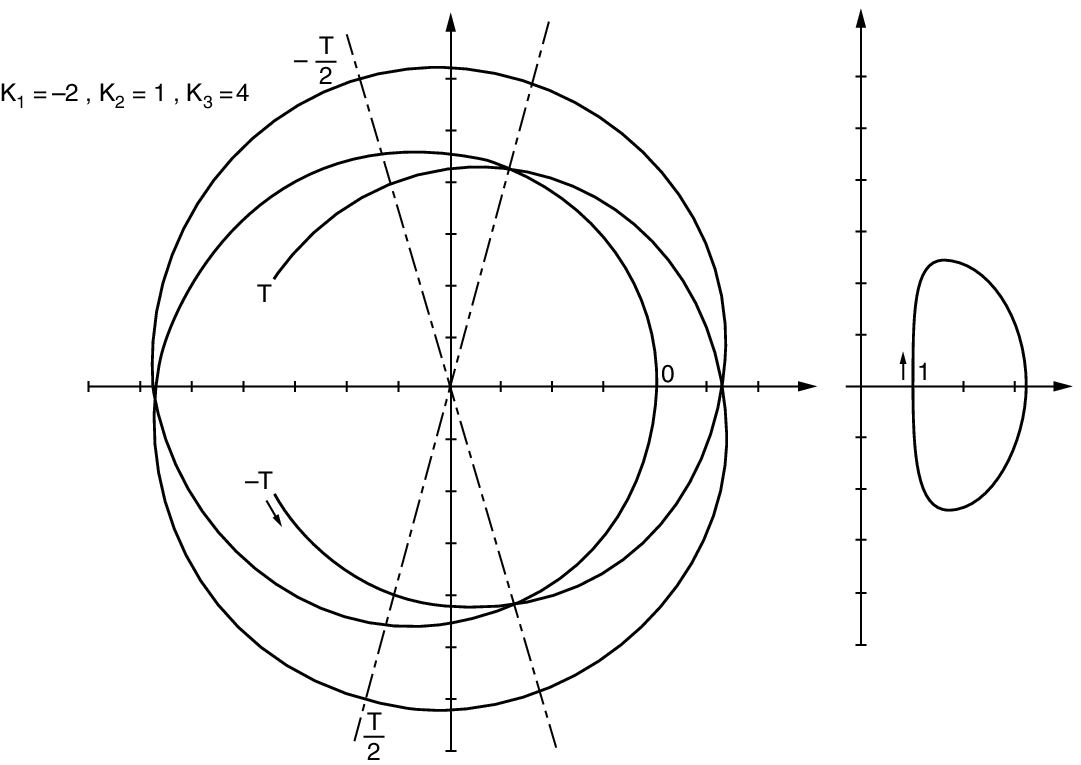}}
 \caption{Absolute motion $z_{1}(t)$ (left) and rotating motion $z_{1}(t)$
$e^{-i\Delta\varphi\frac{t}{T}}$ (right), $-T < t < T$, for arbitrary strengths and 
initial conditions: $\vec{\kappa}=(-2,1,4)$, $\zeta_{o}=-\frac{3K}{2}$.\newline
Data are: $c=1.726$, $\frac{T}{T_{u}}=0.262$,
$\Delta\varphi=507.9^{\circ}$.\label{fig8}} 
\vspace*{1.5pc}
 \centerline{\includegraphics{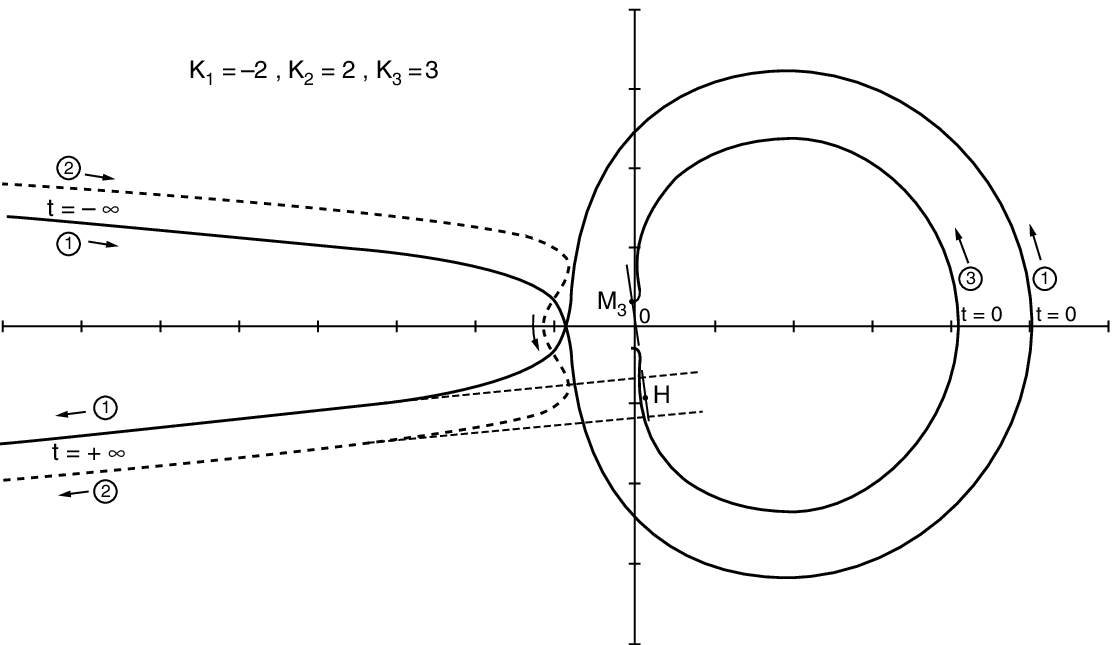}}
 \caption{Absolute motion of the three vortices for an elastic diffusion.
$\vec{\kappa}=(-2,2,3)$, $\zeta_{o}=-\frac{20}{21}K$ $(c=1.170)$.\label{fig9}}
 \vspace*{-25pt}
\end{figure}

Some symmetries may exist in the absolute motion: for instance if the $\zeta$ axis is an
axis of symmetry of the $\zeta$ orbit, then at intervals distant of $\frac{T}{2}$ the
vortices are colinear; if we choose for origin of time an instant of colinearity, the
absolute trajectory of every vortex, when run between 0 and~$T$, possesses an axis of
symmetry (see Figure~\ref{fig8}).


\looseness-1Let us now mention an important result concerning this dynamical
system: since integral (14) is a continuous function of both the strengths
$\frac{\vec{\kappa}}{K}$ and the initial conditions, the angle $\Delta \varphi$ is
generically incommensurable with $2\pi$, i.e. commensurability occurs only for a
set of strengths and initial conditions of zero measure. We conclude that
generically, due to the incommensurability of $\Delta \varphi$ with $2\pi$, the
trajectory of a given vortex completely fills an annulus centered at the
barycentrum (see Landau and Lifchitz Figure~\ref{fig9} Chap.~III).

In the case of a vortex plasma $(K=0)$ and when the barycentrum is at infinity
$(B\ne 0)$, the variation of $z_{j}$ over one period does not depend on $j$ and is
given by
\begin{equation}
\frac{i}{B}[\bar{z}_{j}]_{o}^{T}=\oint\frac{(\zeta^{2}+d\zeta)(\bar{\zeta}^{2}+d\bar{\zeta}-\kappa_{2}\kappa_{3})}
{2\kappa_{1}|\zeta|^{4}\;{\rm Im}(\zeta^{2}+d\zeta)}\D|\zeta|^{2},
\end{equation}
which evaluates to a real quantity. The rotation has become a uniform translation
in the direction normal to the direction of the impulse. The common mean velocity
of the vortices is $[z_{1}]/T$ and, for instance in the case of an initial
equilateral triangle where the absolute motion is a uniform translation, this
velocity evaluates to $\frac{iQ}{2\pi B}$.

Let us conclude all that with a geometrical remark: the relations
\[
\kappa_{1}|Kz_{1}-B|^{2}+ \kappa_{2} \kappa_{3}K|z_{2}-z_{3}|^{2}=sJ,
\]
and
\[
2\pi\frac{\D|z_{2}-z_{3}|^{2}}{\D t}=\frac{2\kappa_{1}s^{2}\;{\rm Im}(\zeta^{2}+d\zeta)}
{|(\zeta+\kappa_{2})(\zeta-\kappa_{3})|^{2}},
\]
show that the extrema of $|z_{j}-\frac{B}{K}|$ occur simultaneously with those of
$|z_{\ell}-z_{m}|$ $(j,\ell,m$ permutation of $1,2,3$) and that they are reached
when the triangle is either flat or isosceles with $M_{j}$ as a summit.

\subsection*{The absolute motion for two or three equal strengths}

An important practical case is that of the invariance of a $\zeta$ orbit under one
of the six permutations of the three elements $(\kappa_{j},z_{j})$. For instance
the permutation (132) acts by:  $\kappa_{2}\to  \kappa_{3}$, $\kappa_{3}\to
\kappa_{2}$ and is seen to leave the set of $\zeta$ orbits invariant provided
$\kappa_{2}=\kappa_{3}$. Since most of the vortices encountered in nature have
equal, or opposite, strengths, we shall consider this particular case in some
detail.

\looseness-1Let us therefore assume $\kappa_{2}=\kappa_{3}$ and consider a $\zeta$
orbit having the origin as a center of symmetry (e.g. the circle $J=0$ on the
Figure~\ref{fig3} assumed continuously deformed so as to admit the origin as a
center of symmetry; note that, on the same figure, the orbit surrounding $T_{1}$
has not the required property). Having chosen such an initial condition $t=0$,
$\zeta=\zeta_{o}$, during the evolution there will happen a time when $\zeta$
evaluates to $-\zeta_{o}$; this time is necessarily equal to the half-period $T/2$
and, since the states ($t=0$, $\zeta=\zeta_{o}$) and $(t=\frac{T}{2}$,
$\zeta=-\zeta_{o}$) are identical from a point of view of initial conditions, we
conclude to the identity of the motions for $0< t < \frac{T}{2}$ and for
$\frac{T}{2}<t<T$; the triangle at $t=\frac{T}{2}$ is deduced from the triangle at
$t=0$ by some fixed rotation around the barycentrum (a change of size and a
translation are excluded because of the invariants) and of course by the exchange
of 2 and 3 $(B=0)$:
\begin{gather*}
\forall\,t\enskip{:}\enskip z_{1}(t)=z_{1}\left(t-\frac{T}{2}\right)e^{i\alpha},\quad z_{2}(t)=z_{3}\left(t-\frac{T}{2}\right)e^{i\alpha},\\
z_{3}(t)=z_{2}\left(t-\frac{T}{2}\right)e^{i\alpha}.
\end{gather*}

By iterating, we see that $2\alpha$ is equal to $\Delta \varphi$ modulo $2\pi$.
Therefore the absolute motions of 2 and 3 are identical, up to a rotation in space
and a translation in time. Every absolute trajectory has two independent axes of
symmetry and therefore, an infinity: they are defined by the barycentrum and the
absolute positions of the vortex when the triangle is either isosceles or flat
(i.e. when $\zeta$ crosses one of its two symmetry axes).

If $K$ is zero, i.e. $\vec{\kappa}=(-2,1,1)$, the conclusions are similar, with a
translation of $L/2$ instead of a rotation, $L$ being the translation after one
period; the initial conditions which have the required symmetry are defined by
$0<c<1$.

Let us now study the absolute motion of three identical vortices. The set of
orbits is invariant under the six permutations, but a given orbit is only
invariant under two or three of them. Two regimes are found:
\begin{enumerate}[(2)]
\item[(1)] $\sqrt[3]{2}<c$. The shortest of the three mutual distances is always the same,
say $M_{2}M_{3}$, the $\zeta$ orbit is invariant by a permutation of 2 and 3 and
the above conclusions apply. The three vortices are colinear at intervals of
$T/2$. Choosing for $t=0$ a colinear configuration, the triangle is isosceles at
$t=\frac{T}{4}+n\frac{T}{2}$, $n\in Z$ and the summit of the triangle is then
always the same vortex $M_{1}$.
\item[(2)] $1<c<\sqrt[3]{2}$, i.e. a vicinity of the equilateral configurations. The $\zeta$
orbit is invariant under any circular permutation and, using quite similar
arguments, we deduce for instance $(B=0)$:
\begin{gather*}
\forall\,t\enskip{:}\enskip
z_{1}(t)=z_{2}\left(t-\frac{T}{3}\right)e^{i\alpha},\quad
z_{2}(t)=z_{3}\left(t-\frac{T}{3}\right)e^{i\alpha},\\[4pt]
z_{3}(t)=z_{1}\left(t-\frac{T}{3}\right)e^{i\alpha},
\end{gather*}
with $3\alpha=\Delta \varphi$ modulo $2\pi$. This time, the three motions are
identical. The vortices are never colinear. Choosing for $t=0$ an isosceles
configuration, we see that the triangle regains the same shape at
$t=n\frac{T}{3}$, successively with the summits $1,2,3$ (or $1,3,2$ depending on
the initial conditions), and a second isosceles shape at
$t=\frac{T}{6}+n\frac{T}{3}$.
\end{enumerate}

The period $T$, that Novikov gave as a hyperelliptic integral, is reducible to an
ordinary elliptic integral (see Appendix~I).

\section{The relative equilibria and the bifurcation set}

In two fundamental papers linking topology to mechanics, Smale (1970) explains how
to split the study of any dynamical system into two simpler problems. He first
defines the integral manifolds as the set of points in the phase space with given
values of the invariants, or better as the quotient of that set by the symmetry
group of the system. Then the first problem is to find the topology of the
integral manifolds of the phase space and more precisely to find the bifurcation
set, i.e. the set of values of the invariants $(E, J)$ for which this topology,
and therefore the nature of the motion, changes. The second problem, which has
been solved above at least in the general case, is the study of the dynamical
systems on the integral manifolds.

Since we do not want to insist on the mathematics, we shall only give the
bifurcation set, i.e., for every value of the strengths, we shall determine the
values of $E$ and $J$ which cause a qualitative change in the absolute motion;
such a research will introduce separating lines in the strengths space.

Our discussion will therefore take place in two different spaces: the space of
strengths (parameter space) and the space of invariants, spaces which we are now
going to describe in more detail.

Due to the homogeneity of the equations of motion, the strengths space may be
represented by its section by a plane $K$ $=$ constant and, in this plane, by a
figure invariant under a rotation of $\frac{2\pi}{3}$ around the point
$\kappa_{1}=\kappa_{2}=\kappa_{3}$. Its dimension is therefore 2 and we shall
represent a point by its polar coordinates:
\begin{equation}
\rho\,\cos\theta=\frac{\sqrt{3}(\kappa_{2}-\kappa_{3})}{2K},\quad
\rho\,\sin\theta=\frac{-2\kappa_{1}+\kappa_{2}+\kappa_{3}}{2K},
\end{equation}
$\theta$ describing any interval of amplitude $\frac{2\pi}{6}$ (see
Figure~\ref{fig1}).

\begin{sidewaystable}
 \tbl{Summary of results.\label{tab1}}
{\tabcolsep12pt\begin{tabular}{@{}lccccccccc@{}} \toprule
&&&\multicolumn{3}{c}{degeneracy}&\\[-5pt]
&&&\multicolumn{3}{c}{\hrulefill}&\\
strengths domain&sign (Q)&sign ($\Delta$)&$c=-\infty$&$c=0$\quad $c=1$&$c=+\infty$&topology\\
\colrule
0&$+$&$-$&0&0&$2,1,2,3$&H M H M H\\
1&$+$&$-$&1&$3,2,1,2$&0&H M H M H\\
1 0&$-$&$-$&$0,1,2,1$&2&2&M E H E M\\
1 1&$-$&$-$&$1,2,1$&2&2, 1&E M H E M\\
1 2&$-$&$-$&$2,1$&2&2, 1, 0&E M H M E\\
1 4 0&$-$&$+$&$0,1$&2&2&M E M\\
1 5 0&$-$&$+$&0&1, 2&2&M E M\\
1 5 1&$-$&$+$&1&2&2, 1&M E M\\
$K=0$, $\kappa_{1}<0<\kappa_{2}<\kappa_{3}$&$-$&$+$&0&2&2&M E M\\
$\Delta=0$ (11 and 351)&$-$&0&1, 1&2&2, 1&E M P M \j{H}\\
$\Delta=0$ (10 and 140)&$-$&0&0, 1, 1&2&2&M E P M \j{H}\\
$\prod (\kappa_{j}+\kappa_{l})=0$ (11 and 12)&$-$&$+$&2, 1&2&2, 1&E M H X \j{H}\\
id. (10 and 11)&$-$&$+$&1, 2, 1&2&2&X H E M \j{H}\\
id. (140 and 351)&$-$&$-$&1&2&2&X M\\
$Q=0$ (1 and 12)&0&$-$&0&2, 1, 2, 4&1&M H M\\
point B $(-5,4,4)$&$-$&0&0, 1&2&2&M Q M\\
point A $(-1,1,1)$&$-$&$-$&2, 2&3&2&X H X\\
point C $(-2,\sqrt{3},2)$&$-$&0&1, 1&2&2&X P M\\
\botrule
\end{tabular}}
\begin{tabnote}
The strengths domains are defined in Figure~\ref{fig1}. Under the heading
``degeneracy'' are listed by  interval of $c$ the number of different states, i.e.
of $\zeta$ orbits, having the same value of $c$; for instance, in the domain 11,
the limiting values of $c$ are $-\infty,0,+\infty$ (at the points $M_{j}$), 1 (at
the triangles), $c_{1}, c_{2},c_{3}$ (at the colinear relative equilibria), they
define 6 intervals, hence a sequence of 6 degeneracy numbers; when $Q$ is zero,
$e^{4\pi E/K^{2}}$ is used in place of $c$ for the classification.

\smallskip
\quad The column ``topology'' lists the sequence of the nature of the real
remarkable $\zeta$ points along the real axis: $M$ stands for a point $M_{j}$ (we
omit $M_{1}$ at infinity), $E, H, P$ for elliptic, hyperbolic, parabolic, $X$ for
an $M$ coinciding with $E$ or $H$ and $Q$ for the only higher-order point we
found. The information on the nature of the triangles is contained in the column
sign $(Q)$. All this is sufficient to draw every $\zeta$ plane.
\end{tabnote}
\end{sidewaystable}

For given values of the strengths, the space of invariants is a priori
bidimensional, since the center of vorticity is not a relevant invariant except
when $K$ is zero. Let us compare this space to the space of initial conditions. As
we have seen, an initial condition is an orbit in the $\zeta$ plane, since two
sets of absolute positions $z_{i}$ whose $\zeta$ values belong to the same $\zeta$
orbit evolve in the same absolute motion, up to a translation in time and a
translation, rotation and scale change in space. On a given orbit, $c$ is
constant, but, inversely, the equation $c=\hbox{cst}$ represents a finite number (between 
0 and 4, see Table~\ref{tab1}) of orbits. From this fact, we draw two conclusions:
first, the space of invariants $(E, J)$ is in fact of dimension one, two points
being identified if they lead to the same value of
$c=\frac{J}{Q}e^{4\pi\frac{E}{Q}}$; second, an initial condition is characterized
by, and therefore equivalent to, a value of the invariant $c$ plus an index of
region in the $\zeta$ plane (or of sheet in the phase space). We can therefore
identify the space of initial conditions to the product of the one-dimensional
space of invariants by the finite set of the indices of region. Accordingly, the
most precise graphical representation will be the orbits of the $\zeta$ plane, but
we shall also use for simplicity a plane $(E, J)$ or an axis $c$ with some
handwritten information on it to take into account the integer index.

Let us now proceed to the determination of the bifurcation set. We exclude the set
of collisions, represented in the $\zeta$ plane by three points of affixes
$\infty$, $\kappa_{3},-\kappa_{2}$ where $c$ evaluates to $+0$, $-0$, $+\infty$ or
$-\infty$ depending on the signs of $J, Q$ and the strengths; these limiting
values of $c$ will therefore belong to the bifurcation set.

The nature of the motion (at least its topological nature since it is always
generically biperiodic) changes when there is a stationary point on the $\zeta$
orbit; such points, where the velocity of $\zeta$ vanishes, are given by the
solutions of the complex equation
\begin{equation}
\bar{\zeta}(\zeta^{2}+d\zeta-Q)-sK(\zeta+d)=0,
\end{equation}
equivalent to
\[
\frac{K\bar{z}_{1}-\bar{B}}{z_{2}-z_{3}}+
\frac{K\bar{z}_{2}-\bar{B}}{z_{3}-z_{1}}+
\frac{K\bar{z}_{3}-\bar{B}}{z_{1}-z_{2}}=0.
\]

These points are also the critical points of the function $(\xi,\eta)\to c$ and
they give all the relative equilibrium configurations of the three vortices, i.e.
the states for which the system moves generally like a rigid body. The solutions
of (17) are:
\begin{enumerate}[(b)]
\item[(a)] one or three colinear configurations, defined by the real zeros
of $P(\zeta)\equiv \zeta^{3}+d\zeta^{2}-(sK+Q)\zeta-sKd$ (points
$P_{1},P_{2},P_{3}$ of Figures~\ref{fig2} to \ref{fig6}),

\item[(b)] two equilateral triangles, defined by the zeros of $\zeta^{2}+d\zeta+sK-Q$, i.e.
$\zeta=\frac{-d\pm is\sqrt{3}}{2}$ (points $T_{1}$ and $T_{2}$ of
Figures~\ref{fig2} to \ref{fig6}); these stationary triangles, already known to
Kelvin for identical vortices, therefore exist for vortices of any strengths.
\item[(c)] When $Q=0$, the isolated zero $\zeta=-d$ of $P$ and every point of the
circle $|\zeta|^{2}=sK$ on which lie the summits of the equilateral triangles and
the two other real zeros of $P$. On this circle $J$ is equal to zero.
\end{enumerate}

The corresponding absolute motions will be described later. It is interesting to
remark that, except for some values of the strengths like for instance $Q=0$, the
relative equilibria of the three vortex system are qualitatively the same than
those of the three body problem of celestial mechanics, where there are two
equilateral triangles and three colinear configurations; in fact, a simple
geometric reasoning shows the same qualitative composition of the set of relative
equilibria for every planar three body motion ruled by a two body central
interaction.

The study of the stability of the relative equilibria is given in Appendix~II and
only three different behaviours are found (in celestial mechanics, only two cases
arise: the three colinear configurations are always unstable, the two triangles
are stable for $(m_{1}+m_{2}+m_{3})^{2}-27(m_{1}m_{2}+m_{2}m_{3}+m_{3}m_{1})>0$,
unstable otherwise), depending on the signs of two polynoms of the strengths of
even degree:
\begin{enumerate}[$\Delta > 0$, $Q < 0$:]
\item[$\Delta > 0$, $Q < 0$:] unstable triangles, one stable aligned configuration,
\item[$\Delta < 0$, $Q < 0$:] unstable triangles, two stable aligned configurations, one \hbox{unstable},
\item[$\Delta < 0$, $Q > 0$:] stable triangles, three unstable aligned configurations,
\end{enumerate}
with
\[
\Delta=-32\sum_{j\ne l}\kappa_{j}^{3}\kappa_{l}-61 \sum_{j<
l}\kappa_{j}^{2}\kappa_{l}^{2}-118\kappa_{1}\kappa_{2}\kappa_{3}\;K.
\]

The important role played by $Q$ can easily be understood if we notice that the
angular momentum of the system
$\sum\kappa_{j}(x_{j}\frac{\D y_{j}}{\D t}-y_{j}\frac{\D x_{j}}{\D t})$ is precisely
$\frac{Q}{2\pi}$. Like for the planar three body problem (see Smale~II), we are
going to see that its sign is a basic element of classification of the topology of
the phase space and that very special behaviours of the motion occur when it
vanishes.

\begin{figure}
 \centerline{\includegraphics{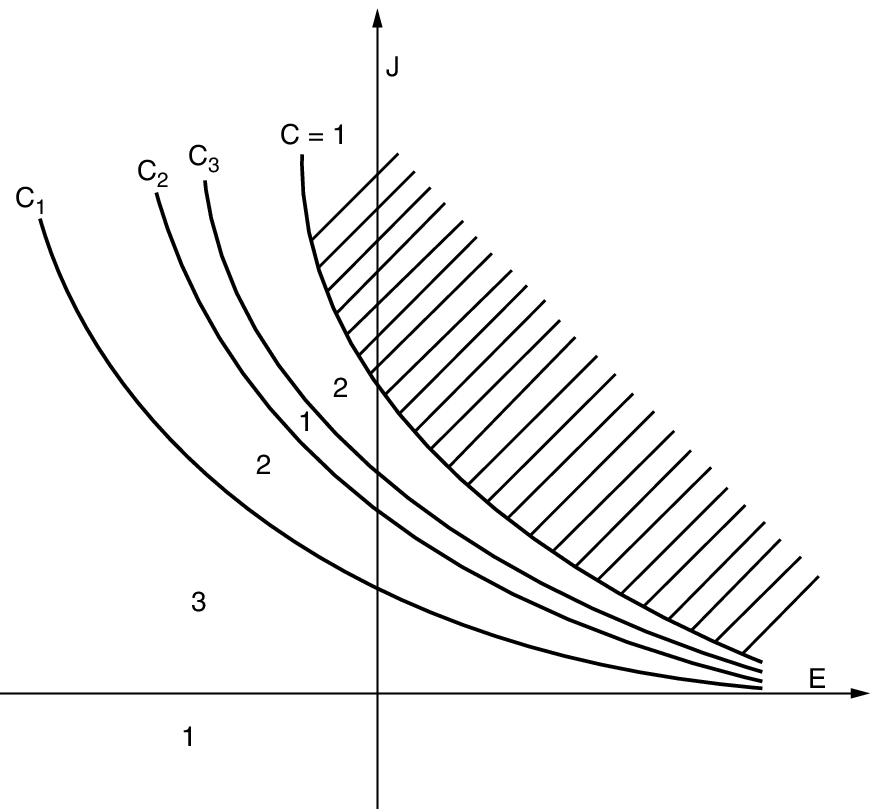}}
 \caption{Bifurcation set in the domain 1 $(Q>0$, $\Delta<0$), not to scale. The
numbers are the number of periodic solutions in each region.\label{fig10}}
\end{figure}

>From (7) we see that there is no other $\zeta$ orbit where the nature of the
motion changes. For given values of the strengths, the bifurcation set is
therefore the union, in the $\zeta$ plane, of the three points $\infty$,
$\kappa_{3},-\kappa_{2}$ and the three or five orbits going through the stationary
points (see Figures~\ref{fig2} to \ref{fig4}, where some other ordinary orbits
have been added). Note that the circle
$|\zeta|^{2}=-\frac{\kappa_{2}\kappa_{3}K}{\kappa_{1}}$ on which $J$ is zero does
not belong to the bifurcation set, except when $Q$ is zero as we shall see.
However, when we represent the bifurcation set in the $(E,J)$ plane, the line
$J=0$ seems to belong to it for it is for some values of the strengths the
limiting curve $c\to 0$ associated with two vortices at the same location;
Figure~\ref{fig10} shows the same bifurcation set as Figure~\ref{fig3}, but
represented in the plane $(E,J)$.

\looseness-1Special behaviour for the motion will be obtained when two or more
elements of the bifurcation set come into coincidence. This will give us
separating lines in the strengths plane. The only conditions of coincidence
between the solutions of (17) and the set (d) of collisions $(\zeta=\infty$,
$\kappa_{3},-\kappa_{2})$ are:
\begin{enumerate}[(ii)]
\item[(i)] coincidence of (a) and (d): $\prod_{j\ne l}(\kappa_{j}+\kappa_{l})=0$,
\item[(ii)] coincidence of two zeros of $P\,{:}\,\Delta=0$.
\end{enumerate}

We must add to the above mentioned lines the three lines given by
$\kappa_{1}\kappa_{2}\kappa_{3}=0$ which are forbidden.
$\prod(\kappa_{j}+\kappa_{l})$ is represented by three lines, $Q=0$ by the circle
$\rho=1$ and $\Delta=0$ by a quartic of equation:
\[
\rho^{4}+8\rho^{3}\sin 3\theta+18\rho^{2}-27=0,
\]
the form of which reflects the ternary symmetry.

The intersections of these lines define three distinct remarkable points that we
shall study explicitly:
\begin{eqnarray*}
&&\hbox{point A: $\rho=2$},\quad \hbox{$\ds 3\theta=\frac{3\pi}{2}$ (strengths $-1,1,1$)},\\[4pt]
&&\hbox{point B: $\rho=3$},\quad \hbox{$\ds 3\theta=\frac{3\pi}{2}$ (strengths $-5,4,4$)},\\[4pt]
&&\hbox{point C: $\rho=\sqrt{5}$},\quad \hbox{$\ds
\sin\;3\theta=\frac{-11\sqrt{5}}{25}$ (strengths $-2,\sqrt{3},2$).}
\end{eqnarray*}

The points A, B and some numerical values of the strengths correspond to cases of
integrability for the period (see Appendix~I).

These lines define different domains in the strengths space, which has been
represented on Figure~\ref{fig1}. When we stay in one of these domains, the
topology of the bifurcation set does not change, and Table~\ref{tab1} summarizes
the results concerning the bifurcation set for all domains of the strengths space.

The classification for the bifurcation set is mainly based on the signs of $Q$ and
$\Delta$, i.e. on the number and stability of the relative equilibria; among the
three quantities, i.e. arithmetic, geometric and harmonic means of the strengths,
whose signs were proposed by Aref as a basis for a classification, only the third
one $\frac{Q}{3\kappa_{1}\kappa_{2}\kappa_{3}}$ is relevant, although the factor
$\kappa_{1}\kappa_{2}\kappa_{3}$ prevents it from being invariant under a
strengths reversal, an operation equivalent to a time reversal which must leave
invariant any candidate to a classification; moreover, for any $N$, the formula
$\sum'\kappa_{j}\kappa_{l}=0$, not $\sum\frac{1}{\kappa_{j}}=0$, expresses the
associated physical property of invariance of the energy under a change of length.

In the next part, we describe the behaviour of the three vortex system when it is
nongeneric, i.e. when either the strengths are on the boundaries of the domains
represented on Figure~\ref{fig1} or when the initial conditions are those of a relative
equilibrium. We shall proceed by studying first the relative equilibria associated with
nonsingular strengths, then the strengths on the separating lines of the strengths space
and finally the three particular points A, B, C of the strengths space.

\section{Absolute motions for nongeneric strengths\\ or initial conditions}

\subsection*{Absolute motions for the ordinary relative equilibria}

Ordinary means that we exclude $\zeta$ points which are the coincidence of two elements
of the bifurcation set: this is equivalent to $J$ nonzero and $\zeta$ not a multiple zero
of $P$.

For an initial condition in the vicinity of such a $\zeta$ point, every point obeys the
general motion, except those lying on the two curves intersecting at an unstable $\zeta$
point, in which case the motion is asymptotic to the motion at the stationary point,
motion which we are going to determine:

\medskip\noindent(a) $K\ne 0$. From the time variation law
\[
2\pi\frac{\D |Kz_{1}-B|^{2}}{\D t}=\frac{-2s^{2}{\rm
Im}(\zeta^{2}+d\zeta)}{|(\zeta+\kappa_{2})(\zeta-\kappa_{3})|^{2}},
\]
it follows that the distance from each point to the barycentrum remains constant.
The absolute motion is therefore a solid rotation around the barycentrum;
moreover, using
\[
\sum_{j=1}^{N}\kappa_{j}z_{j}\frac{\D\bar{z}_{j}}{\D t}=\frac{Q}{2\pi i},
\]
whose imaginary part yields
\[
\sum_{j=1}^{N}\kappa_{j}|z_{j}|^{2}\frac{\D\arg z_{j}}{\D t}=\frac{Q}{2\pi},
\]
we see that the common angular velocity of every vortex is independent of time and
remains equal to $\frac{Q}{2\pi I_{o}}$, where $I_{o}$ is the inertia momentum
relative to the barycentrum
\[
I_{o}=\sum \kappa_{j}\left|z_{j}-\frac{B}{K}\right|^{2}=\frac{J}{K},
\]
and $\frac{Q}{2\pi}$ is the angular momentum of the system.

The period of the uniform rotation is therefore
\[
T_{u}=\frac{4\pi^{2}J}{QK},
\]
a formula valid for any number of vortices when there exists a solid rotation.

A particular case is $Q=0$, for which the only isolated relative \hbox{equilibrium} is
$\zeta=-d$, i.e.
$\kappa_{2}\kappa_{3}z_{1}+\kappa_{3}\kappa_{1}z_{2}+\kappa_{1}\kappa_{2}z_{3}=0$; all
the vortices remain at rest. This situation is of course unstable (free vortices cannot
have a stable rest position since the complex velocity is a nonconstant meromorphic
function of the affixes), and the small motions have for pulsation
$\omega=\pm\frac{9i\kappa_{1}\kappa_{2}\kappa_{3}}{2\pi J}$

\medskip\noindent (b) $K=0$. The three stationary points are ordinary.

$\zeta=0$, stable: the impulse $B$ is zero and the affixes verify:
\[
\frac{z_{2}-z_{3}}{\kappa_{1}}=\frac{z_{3}-z_{1}}{\kappa_{2}}=
\frac{z_{1}-z_{2}}{\kappa_{3}}.
\]

The inertia momentum is nonzero and is the same at every point, and the absolute motion
is a uniform solid rotation of period
\[
T=\frac{4\pi^{2}I}{Q},
\]
equal to that of the small motions.

At the two unstable equilateral triangles, $B$ is nonzero and the absolute motion is a
uniform translation of velocity
\[
\forall\,j\enskip{:}\enskip\frac{\D z_{j}}{\D t}=\frac{iQ}{2\pi \bar{B}},
\]
orthogonal to the impulse.

\subsection*{Absolute motions on the boundaries of the strengths domains}

These lines are:
$Q(\kappa_{2}+\kappa_{3})(\kappa_{3}+\kappa_{1})(\kappa_{1}+\kappa_{2})\Delta=0$.

\subsection*{$Q=0$. Triple collision in a finite time, expanding motion}

In addition to the already studied isolated stationary point, there exists a
circle of stationary $\zeta$ points (see Figure~\ref{fig5}), on which $J$ is zero,
On this circle lie the two summits of the equilateral triangles and the two other
zeros of $P$. These four points are the points of contact of the circle $J=0$ with
the set of $\zeta$ trajectories whose equation is now:
\[
\left|\frac{\zeta-\kappa_{3}}{s}\right|^{-2\frac{\kappa_{1}\kappa_{2}}{K_{2}}}
\left|\frac{\zeta+\kappa_{2}}{s}\right|^{-2\frac{\kappa_{1}\kappa_{3}}{K_{2}}}=e^{\frac{4\pi
E}{K^{2}}}.
\]

We shall first study the absolute motion when the $\zeta$ point lies on the circle
$J=0$, then examine its neighborhood and finally deduce the bifurcation~set.

Let us assume $\zeta$ on the circle $J=0$. This circle is no longer a trajectory and the
$\zeta$ point stays at rest. The two conditions $Q\,{=}\,0$, $J\,{=}\,0$ express that
both the energy and the inertia momentum relative to the barycentrum are invariant under
a change of length, and therefore nothing prevents the vortices from going to infinity or
to zero; we see below that both cases are possible. The absolute motion is ruled by
Equation~(6) which implies that $\frac{d\rho_{1}^{2}}{dt}$ and
$\rho_{1}^{2}\frac{d\varphi_{1}}{dt}$ are constant in time ($\rho_{1},\varphi_{1}$ are
polar coordinates of $M_{1}$, $B$ is chosen zero). Since the shape of the triangle is
conserved, Equation~(6) integrates as~in
\begin{eqnarray*}
\forall\,j&=&1,2,3\enskip{:}\enskip z_{j}=z_{j,o}\left(1-\frac{t}{t_{c}}\right)^{1/2-i\omega t_{c}},\\
\hbox{i.e.}\quad
\sqrt{1-\frac{t}{t_{c}}}&=&\frac{\rho_{j}}{\rho_{j,o}}=\exp\{(\varphi_{j}-\varphi_{j,o})/(-2\omega
t_{c})\},
\end{eqnarray*}
where the characteristic time $t_{c}$ and the initial angular velocity $\omega$
are defined~by
\[
-2\omega+\frac{i}{t_{c}}=\frac{-s^{2}}{\pi
K|z_{1,0}|^{2}}\frac{\zeta^{2}+d\zeta}{\zeta^{2}+d\zeta-\kappa_{2}\kappa_{3}}=\frac{1}{3\pi}{\sum}'_{j,\ell}\dots
 \frac{\kappa_{\ell}}{(\bar{z}_{j}(z_{\ell}-z_{j}))_{o}}.
\]
\looseness-1$\omega$ never vanishes and has the sign of $K$. $\frac{1}{t_{c}}$
vanishes and changes sign when $\zeta$ is one of the four points already mentioned
where the circle is tangent to the set of $\zeta$ curves. We conclude that, for a
$\zeta$ point of the circle $J=0$ distinct of these four points, every vortex runs
a logarithmic spiral whose pole is the barycentrum, the shape of the triangle
remains constant and, depending on the sign of $t_{c}$, the triangle either
expands to infinity in an infinite time or collapses on the barycentrum in a
finite time $t_{c}$. At the time of this triple collision all the denominators of
the equations of motion (1) simultaneously vanish like $(t_{c}-t)^{1/2}$. After
the collision, the system is made of a single motionless vortex of strength $K$
located at what was the center of vorticity.

When $\frac{1}{t_{c}}$ vanishes, the spiral motion degenerates into a uniform
circular motion whose period $T=\frac{2\pi}{\omega}$ can also be written as
\[
T=\frac{4\pi^{2}}{6K}(|z_{2}-z_{3}|^{2}+|z_{3}-z_{1}|^{2}+|z_{1}-z_{2}|^{2}),
\]
for the two aligned configurations and twice the same expression for the
equilateral triangles.

Let us now examine the motion elsewhere in the $\zeta$ plane $(J\ne 0)$.
Figure~\ref{fig5} shows that two generic situations exist depending on whether the
$\zeta$ orbit intersects or not the circle $J=0$ of singular points; the limiting
$\zeta$ orbits, which belong to the bifurcation set, are $E=0$ (tangent at
$T_{1},T_{2}$) and $E=E(P_{3})$. For an energy $E$ outside the interval
$[E(P_{3}),0]$, the motion is the usual biperiodic one.

Inside the energy interval $]E(P_{3}),0[$, striped on Figure~\ref{fig5}, every $\zeta$
trajectory stops on the circle (note that it cannot cross it) and, since $J$ is the
product of $|z_{2}-z_{3}|^{2}$ by a function of $\zeta$ vanishing on the circle, every
$\zeta$ point having a nonzero $J$ and the energy of a curve intersecting the circle
yields an expanding motion in which the trajectory of every vortex is asymptotic to a
logarithmic spiral going to infinity. No motion exists which is asymptotic to the triple
collision in a finite time: therefore the points of the half circonference (from $P_{1}$
to $T_{1}$ and from $P_{3}$ to $T_{2}$) where such a collision exists are repulsive
points, while the other half is made of attractive points.

\begin{figure}
 \centerline{\includegraphics[scale=1.05]{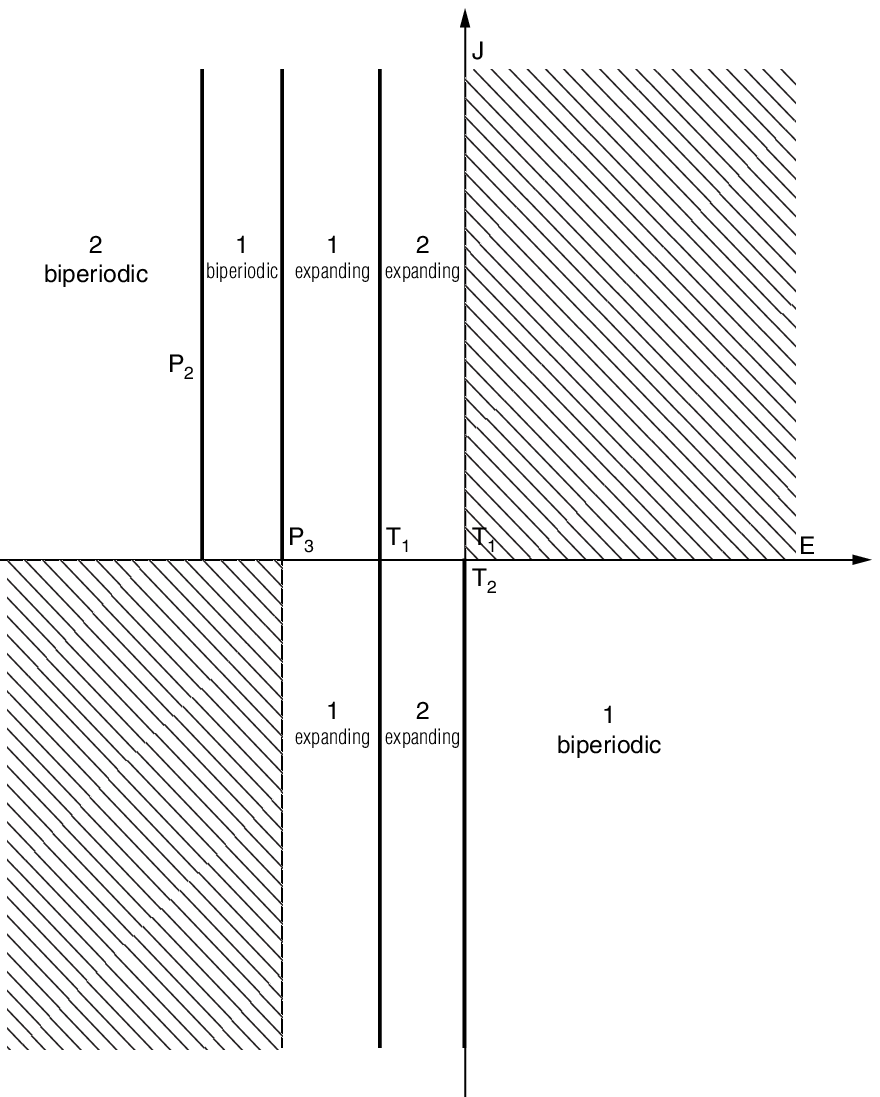}}
 \caption{Bifurcation set for $Q=0$, not to scale.\label{fig11}}
 \end{figure}

Finally, the bifurcation set, which we have also represented in the $(E,J)$ plane
on Figure~\ref{fig11}, is the union of the following lines of the $\zeta$ plane:
the~set of collisions $(\zeta=\infty$, $-\kappa_{2},\kappa_{3}$), the circle $J=0$
(not an orbit), the orbit going through $P_{2}$, the orbit $E=0$ tangent to the
circle at $T_{1},T_{2}$, the orbit $E=E(P_{3})$ tangent to the circle at $P_{3}$
and that portion of the orbit $E=E(P_{1})$ which is interior to the circle $J=0$.

\subsection*{$\Delta=0$}

Two of the three colinear relative equilibria coincide except at the three points
of retrogression $\rho=3$ of the quartic $\Delta=0$ (strengths $-5,4,4$) where the
three of them coincide and at the six points $\rho=\sqrt{5}$ (strengths
$-2,\sqrt{3},2$) where two of the strengths are opposite (see next case).

For the absolute motion, there is nothing qualitatively changed: in the vicinity
of the double point $\zeta_{o}$, the orbits are equivalent to the cubics of
equation
\[
c-c_{o}+A_{2}\eta^{2}+A_{3}(\xi-\xi_{o})^{3}=0,
\]
and the period, which behaves like $J\int\frac{\D\xi}{\eta}$, diverges like
$|c-c_{o}|^{-1/16}$; the absolute motion is still biperiodic, although the
triangle spends quite a long time in nearly aligned states. On the orbit $c=c_{o}$
passing through the unstable $\zeta_{o}$ point, the absolute motion is asymptotic
to the usual circular motion, the time law being different: $\xi-\xi_{o}\sim
t^{-2}$, $\eta\sim t^{-3}$.

\subsection*{$\Pi(\kappa_{j}+\kappa_{\ell})=0$. Elastic diffusion}

At least two vortices have opposite strengths. To fix the ideas, let us assume
$\kappa_{1}+\kappa_{2}=0$. The case $\vec{\kappa}=(-1,1,1)$ will be described in
next section. The point $\zeta=\kappa_{3}$, which lies on the circle $J=0$, is
singular in the sense that for every $c$ value there exists a $\zeta$ orbit
passing through $\kappa_{3}$, which forbids a reduced periodic motion.

Therefore two generic situations exist, as shown on Figure~\ref{fig6}: outside of
the striped domain which contains $J=0$, the absolute motion is biperiodic. Inside
this domain, for every $c$ value there is one and only one $\zeta$ orbit and the
$\zeta$ point is attracted by $\kappa_{3}$ in an infinite time: it tends to
$\kappa_{3}$ normally to the real axis following a law
\[
i(\zeta-\kappa_{3})t\to \frac{2\pi s}{\kappa_{1}}e^{-4\pi E/Q}.
\]

In the absolute space, the point $M_{3}$ stops and the two others go to infinity
together with a motion asymptotic to a uniform translation in the direction normal
to the line joining the final position of $M_{3}$ to the barycentrum, exactly as
if they were alone and obeyed the motion of two vortices of opposite strengths: at
the limit, the translation velocity is $\frac{\sqrt{-Q}}{2\pi}e^{2\pi E/Q}$ and
the mutual distance is $e^{-2\pi E/Q}$.

This creation of a doublet can also be interpreted in terms of the elastic diffusion of
the doublet by the third vortex: if we take for initial condition $\zeta$ near to $K_{3}$
with $\eta/K$ positive, the vortices 1 and 2 are initially a doublet moving towards the
third vortex; then the three mutual distances become the same order of magnitude, i.e.
there is interaction; finally, the doublet emerges and goes to infinity away from the
third vortex with unchanged mutual distance and velocity, but in a different direction.
This is exactly a process of elastic diffusion. Figure~\ref{fig9} shows typical absolute
motions. We define the scattering angle $\Delta(\phi)$ by the total variation along the
trajectory of $\arg(z_{1})$ ($B$ is chosen zero) and a dimensionless impact parameter by
$\frac{\overline{OH}}{\overline{OM_{3}}}$ (Figure~\ref{fig9}) where $H$ is the projection
of the middle of $M_{1}M_{2}$ on the line defined by the barycentrum and the final
position of $M_{3}$:
\begin{eqnarray*}
\frac{\overline{OH}}{\overline{OM_{3}}}&=&\frac{{\rm Re}(\bar{z}_{3}\frac{z_{1}+z_{2}}{2})}{\bar{z}_{3}z_{3}}\\[4pt]
&=&\frac{{\rm
Re}((\vec{\zeta}_{3}-\kappa_{3})(\kappa_{3}^{2}-\kappa_{1}\zeta+s\zeta))}{-2\kappa_{1}|\zeta-\kappa_{3}|^{2}}\sim
\frac{s\kappa_{3}}{-2r\kappa_{1}}+\frac{s-\kappa_{1}}{-2\kappa_{1}}=\frac{c+1}{2},
\end{eqnarray*}
where $r$ is the radius of curvature in $\kappa_{3}$:
\[
r=\lim\frac{\eta^{2}}{2(\xi-\kappa_{3})}=\frac{-s\kappa_{3}}{s+\kappa_{1}c}.
\]
The impact parameter is then our dimensionless invariant $c$, up to a linear
transformation. As to the scattering angle, it can be computed from an integral taken
along the $\zeta$ orbit:
\begin{eqnarray*}
\Delta \varphi_{1}=[\arg\, z_{1}]_{t=-\infty}^{+\infty}=\oint -sK\frac{{\rm 
Re}\left(\frac{\zeta^{2}+d\zeta}{\zeta^{2}+d\zeta-\kappa_{2}\kappa_{3}}\right)}
 {{\rm
Re}\left(\frac{\zeta[\vec{\zeta}(\zeta^{2}+d\zeta-Q)-sK(\zeta+d)]}
{\zeta^{2}+d\zeta-\kappa_{2}\kappa_{3}}\right)}\D(\arg\;\zeta).
\end{eqnarray*}
This integral can be carried out exactly for $c=0$, when the $\zeta$ orbit is a
circle:
\begin{eqnarray*}
\Delta \varphi_{1}&=&\int_{o}^{2\pi}\frac{2K+d}{2(K+d)}\left(1-\frac{Kd}{2K^{2}+2Kd+d^{2}+2K(K+d)\cos\theta}\right)\D\theta\\[4pt] 
&=&\left\{\begin{array}{@{}l@{\quad}l@{}} 
\pi&{\rm if}\ |\kappa_{1}|>|K|\\[7pt]
\ds \left(1-2\frac{K}{\kappa_{1}}\right)\pi&{\rm if}\ |\kappa_{1}|<|K|.\\
\end{array}\right.
\end{eqnarray*}

The bifurcation set for this case is made, in the $\zeta$ plane, of the following
lines: the set of collisions $(\infty,\kappa_{3},-\kappa_{2})$, the boundary
between the two regimes, the point $P_{1}$ and the orbit $c=c(P_{2})$.

 \enlargethispage{12pt}

\subsection*{The three particular $\vec{\kappa}$ points}

(a) Point A (strengths $-1,1,1$). Direct and exchange diffusion.

The $\zeta$ plane looks like the one Figure~\ref{fig6} assumed continuously
deformed so as to admit the origin as a center of symmetry: $M_{3}=P_{1}$,
symmetric of $M_{2}$, and $P_{2}=0$.

Aref (1979) has extensively studied the motion and we shall only briefly summarize it.
The results obtained for $\kappa_{1}+\kappa_{2}=0$ still apply, but a new physical
situation arises from the existence of $\zeta$ orbits which go from $\zeta=\pm K$ to
$\zeta=\mp K$ (these are all the orbits which cross the segment $T_{1}T_{2}$): the motion
is then an exchange scattering in which the incident pair (1 2) is different from the
outgoing doublet (1 3).

To sum up, three generic situations exist, depending on the initial conditions:
\begin{itemlist}
\item $|\frac{\zeta}{K}\pm 1|>2$: in uniformly rotating axes, the absolute motion is
periodic.
\item $|\frac{\zeta}{K}\pm 1|<2$ and $-8<c<1$: exchange scattering;
the integral giving the scattering angle is to be taken from 0 to $\pi$ only.
\item $|\frac{\zeta}{K}\pm 1|<2$ and $c$ outside $[-8,1]$: direct scattering as previously described.
\end{itemlist}

The reduced forms of all the elliptic integrals giving the period and the
scattering angle are gathered in Appendix~I.

A particular case of exchange scattering is easy to solve and will give an idea of
the motion:

For $c=0$, the $\zeta$ orbit is a half-circle and, computing
$\frac{\D^{2}z_{1}}{\D t^{2}}$ by deriving Equation~(6), we find zero, which means a
uniform linear motion $M_{1}$, hence a scattering angle of $\pi$. The motion of
$\zeta=Ke^{i\theta}$ is ruled by $\frac{\D\theta}{\D t}=\frac{v}{a}\sin^{2}\theta$
which integrates as $\zeta=K\frac{ia-vt}{\sqrt{a^{2}+v^{2}t^{2}}}$ where we have
noted $a=2e^{2\pi E/K^{2}}$, $v=\frac{K}{\pi a}$.

The origin of time being chosen when $M_{1}$ is the summit of an isosceles
rectangle triangle, the absolute motions take place on three parallel
straight~lines:
\[
\left\{\begin{array}{@{}l@{}}
\ds z_{1}=a+i\;v\;t\\[4pt]
\ds z_{2}=\frac{a+i\;v\;t}{2}+\frac{i}{2}\sqrt{a^{2}+v^{2}\;t^{2}}\\[10pt]
\ds z_{3}=\frac{a+i\;v\;t}{2}-\frac{i}{2}\sqrt{a^{2}+v^{2}\;t^{2}}.\\
\end{array}\right.
\]

The exchange scattering process is clearly seen on the above equations, and every
other exchange scattering motion can be thought of as a continuous deformation of
this one.

 \enlargethispage{12pt}

The bifurcation set is made of the set of collisions and of the boundaries of the
domains limiting the three generic situations; note that the circle
$J=0(|\zeta|=|K|)$ does not belong to it. This set is simple enough to be
represented without any ambiguity by a c axis with the number and type of
solutions in each interval:

\bigskip
\centerline{\includegraphics{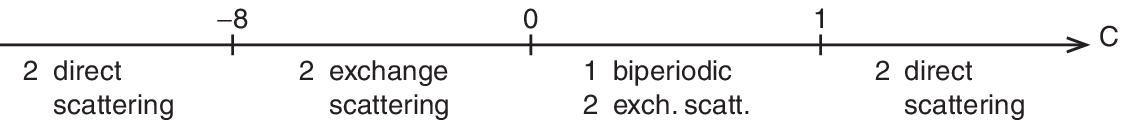}}

\medskip\noindent (b) Point B summit of the quartic (strengths $-5,4,4$).

Nothing special happens to the motion. At the triple point $\zeta=0$, the vortex of
strength $\frac{-5K}{3}$ is motionless as it coincides with the barycentrum, and the two
other vortices, which are symmetric with regard to the barycentrum, obey the circular
motion. The $\zeta$ orbits around this stable point are equivalent to the quartics of
equation
\[
2^{-4/3}c+1-\frac{15}{8}\left(\frac{\eta}{K}\right)^{2}-\frac{45}{256}
\left(\frac{\xi}{K}\right)^{4}=0,
\]
and they correspond to ordinary motions; however the period, which still behaves
like $J\int \frac{\D\xi}{\eta}$, diverges as $(c+2^{4/3})^{-1/4}$.

\medskip\noindent
(c) Point C (strengths $-2, \sqrt{3}, 2$) conjunction of $\Delta=0$ and
$\Pi(\kappa_{j}+\kappa_{\ell})=0$. This point has no new properties: the
previously studied singularities only add without interfering.

\section{Volume of the phase space}

The Hamiltonian system has an invariant element of volume of the phase space equal
to
\[
\D v=\kappa_{1}\kappa_{2}\kappa_{3}\D x_{1}\wedge \D y_{1}\wedge \D x_{2} \wedge \D y_{2}
\wedge \D x_{3}\wedge \D y_{3}.
\]

Since there exist four real invariants $E, J, X, Y$ $(X+iY=B)$, we want the
density of states $\frac{\D v}{\D E\;\D J\;\D X\;\D Y}$ after integration over two
independent variables of the phase space.

By using the two successive changes of variables $(z_{1},z_{2},z_{3})\to
(z_{2}-z_{3},\zeta,B)$ and
\[
(\zeta,\bar{\zeta},z_{2}-z_{3},\overline{z_{2}-z_{3}})\to
(E,J,z_{2}-z_{3},\overline{z_{2}-z_{3}}),
\]
whose jacobians are respectively
\[
\frac{D(z_{2}-z_{3},\zeta,B)}{D(z_{1},z_{2},z_{3})}=\frac{-sK}{z_{2}-z_{3}},
\]
and
\[
\frac{D(E,J)}{D(\zeta,\bar{\zeta})}=\frac{i\kappa_{1}^{2}\kappa_{2}\kappa_{3}}{2\pi}|z_{2}-z_{3}|^{2}\frac{{\rm
Im} (\zeta^{2}+d\zeta)}{|(\zeta-\kappa_{3})(\zeta+\kappa_{2})|^{2}},
\]
we obtain
\begin{eqnarray*}
\D v&=&\frac{\pi}{\kappa_{1}s^{2}K^{2}}\frac{|(\zeta-\kappa_{3})^{2}(\zeta+\kappa_{2})^{2}|}{{\rm
Im}(\zeta^{2}+d\zeta)}\\
&& \ \D(x_{2}-x_{3}) \wedge \D(y_{2}-y_{3})\wedge \D E\wedge \D J \wedge \D X\wedge \D Y,
\end{eqnarray*}
and we still have to integrate over $x_{2}-x_{3}$ and $y_{2}-y_{3}$. Since
$|z_{2}-z_{3}|$ moves in time according to
\[
2\pi\frac{\D |z_{2}-z_{3}|^{2}}{\D t}=2\kappa_{1}s^{2}\frac{{\rm
Im}(\zeta^{2}+d\zeta)}{|(\zeta-\kappa_{3})(\zeta+\kappa_{2})|^{2}},
\]
the integration is quite easy to perform and we finally get
\[
\D v=\frac{\pi}{K^{2}}T(E,J) \D E\wedge \D J\wedge \D X\wedge \D Y,
\]
which shows that the density of states is the period of the reduced motion. This
is a well known result of the theory of adiabatic invariants (see e.g. Landau and
Lifchitz), for energy and time are conjugate variables: when $J$ is constant and
$E$ slowly varying, then the product $ET$ is constant since the volume of the
phase space if conserved.

To take into account the fact that the phase space is the union of disconnected
parts, we must sum the above expression over the number (between 0 and 4, see
Table~\ref{tab1}) of different domains associated to given values of $E$ and $J$.
The resulting volume $\Omega(J,E)$ obeys the scaling law:
\[
\Omega(E,J)=\frac{\pi}{K^{2}}\sum_{{\rm
domains}}T(E,J)=\left|\frac{J}{QK^{3}}\right|f\left(\frac{J}{Q}e^{4\pi\frac{E}{Q}}\right).
\]

Onsager (1949) defined the entropy $S$ and the temperature $\tau$ of an assembly
of a large number of interacting vortices:
\[
S=k_{B}\;{\rm Log}\,\Omega,\quad \frac{1}{\tau}=\frac{\D S}{\D E}.
\]

Although it makes no sense to speak of thermodynamics about an integrable system,
there may be some interest for the understanding of the behaviour of a large
number of vortices to examine what the Onsager's theory gives when formally
applied to the three-vortex system.

The main hypothesis made by Onsager is that the total amount of volume
$\int_{-\infty}^{+\infty}\Omega \D E$ available to the system is finite, an hypothesis
equivalent to assume the system confined in a box since the phase space and the
configuration space are the same. Due to the scaling law for $\Omega$, the integral
$\int\Omega \D E$ will be either finite and proportional to $J$ or infinite, depending on
the strengths of the vortices. The first arising question is therefore: when $J$ is kept
constant, is the integral
$\int\Omega(E,J)\D E=\frac{Q}{4\pi}\int_{cQJ>0}\Omega\frac{\D c}{c}$ finite or not, i.e. are
all the singularities of $T$ integrable or not?

The singularities of $T$ are: the set of collisions, the unstable relative
equilibria and, since $J$ is kept nonzero, the limit $e^{4\pi E/Q}\to 0$ in the
domains $\kappa_{1}\kappa_{2}\kappa_{3}K<0$ only.

\medskip\noindent
(a) Two vortices close to each other (vortex $j$ alone): then
\[
T\sim
\frac{4\pi^{2}J}{(K-\kappa_{j})^{2}\kappa_{j}}\left(\frac{Qc}{\kappa_{j}(K-\kappa_{j})}\right)^{-\frac{Q\kappa_{j}}{\kappa_{1}\kappa_{2}\kappa_{3}}}\to
0,
\]
and the singularity $\int\;T$ $\D E=\frac{QJ}{4\pi}\int\frac{T}{J}\frac{\D c}{c}$ is
integrable.

\medskip\noindent
(b) $\zeta$ tends to an unstable $\zeta_{0}$ value, whose $c$ value is $c_{o}$.
The equivalent hyperbola having for equation:
\begin{eqnarray*}
G(\zeta,\bar{\zeta})&\equiv&
\frac{c-c_{o}}{c_{o}}+\frac{J\,\kappa_{1}\kappa_{2}\kappa_{3}}{2Q(\kappa_{1}|\zeta_{0}|^{2}+\kappa_{2}\kappa_{3}K)^{2}}\\[3pt]
&&\times\,[\mu_{o}(\zeta-\zeta_{o})^{2}+2v_{o}|\zeta-\zeta_{o}|^{2}+\bar{\mu}_{o}(\bar{\zeta}-\bar{\zeta}_{o})^{2}]=0,
\end{eqnarray*}
the period is equivalent to
\[
T\sim {\rm cst}\,J\int \frac{\D \bar{\zeta}}{G'_{\zeta}(\zeta,\bar{\zeta})}\sim {\rm cst}\
J\ {\rm Log}|c-c_{o}|,
\]
and the singularity is therefore integrable.

\medskip\noindent
(c) $|\zeta|^{2}\to - \frac{\kappa_{2}\kappa_{3}K}{\kappa_{1}}$ (possible in every
domain, except 0 and 150).

Since $c$ tends to zero with $J$ being kept constant and nonzero, then $e^{4\pi E/Q}$
tends to zero, therefore the period $T=\hbox{cst}$ $e^{-4\pi E/Q}$ has a nonintegrable 
singularity.

In conclusion, the volume of the phase space, with $J$ being constant, is finite
and proportional to $J$ in the domains 0, 150 and $K=0$ and infinite elsewhere.
The unit of time we chose, i.e. $4\pi^{2}\frac{J}{QK}$, is a posteriori convenient
for it is proportional to the volume.

 \enlargethispage{12pt}

Let us now examine the behaviour of the thermodynamical quantities, keeping in mind that
any conclusion is meaningless for three vortices and can only be indicative for a larger
system. For instance in the domain 0 (i.e. $\kappa_{j}K>0$), the function $E_{o}\to
\int_{E<E_{o}}$ $\Omega \D E$ is of course increasing, but the integrand $\Omega$, which is
zero at the edges $E\to \pm \infty$, positive and integrable, has three infinite maxima
at finite values of the energy (Figure~\ref{fig7}). Each of these maxima corresponds to
an unstable relative equilibrium configuration or, in other words, to a point of the
phase space which links two disconnected domains. This feature somehow complicates the
correspondence between energy and temperature and there exists some numerical evidence
(Lundgren and Pointin, 1977, and references herein) of a possible lack of ergodicity
which could come out of a multiple connexity of the phase space. Another interesting
observation is that, when the energy tends to $+\infty$, the ``temperature'' of the three
vortex system tends to a constant, negative value (Fig.~\ref{fig7}), a fact already
noticed for a large number of vortices by Lundgren and Pointin (page~334), C.E. Seyler
(1974), Edwards and Taylor (1974, page 262).

\section*{Conclusion}

In addition to the fact of being an exactly soluble three body problem, the three
vortex system is very interesting in connection with the theory of turbulence.
Unfortunately its number of degrees of freedom is too small to yield a chaotic
behaviour (the threshold for such a behaviour is 3) and this was confirmed by the
results: nonperiodic behaviours are obtained only for very particular values of
the parameters. A four vortex system (see some preliminary results in Conte,
1979), with its 3 independent degrees of freedom and because we do not know about
its integrability, is the really interesting dynamical system to study in order to
have some hints about the integrability of the N vortex system.

\section*{Acknowledgements}

We want to thank Y.~Pomeau for many fruitful discussions which led to the discovery of
the appropriate plane. We also greatly appreciated the formal Reduce-like computer
language AMP (Drouffe, 1976) which helped us to establish the numerous necessary
formulae.

The work is the first chapter of the unpublished Th\`ese d'\'Etat of the first
author. It is an honor and a pleasure for us to dedicate it to Professor Hao
Bailin and to wish him a long life.

 \enlargethispage{12pt}

\pagebreak
\section*{Appendix I}

\section*{Cases of integrability} 

For practical applications, it may be of interest to find which values of the
$\kappa_{j}$'s lead to integrable expressions for the period (12). A first case is
when two strengths are equal: $\kappa_{2}=\kappa_{3}$; then $\xi\eta$ can be
expressed only with $|\zeta|^{2}$, using (11), and the period is a simple integral
in the variable $|\zeta|^{2}$, which can be easily integrated numerically.

Another case is when, the strengths being rational, the algebraic curve (11) is of
genus one or zero (the genus of an algebraic curve of degree $n$ is equal to
$\frac{(n-1)(n-2)}{2}$ minus the number of double points). The only curve of genus
zero is the circle $J=0$ but then the period is given by another non-integrable
expression. If the trajectory is of genus one, the abelian integral expressing the
period can always be reduced to an elliptic integral by a birational
transformation of the coordinates (see e.g. Bateman 1953). For small integer
values of the strengths, there is some chance of finding curves of genus one.

Let us just mention three particular cases.

$\underline{\vec{\kappa}=(1,1,1) K/3}$. This belongs to the first but not to the
second case (degree 6, genus 4 in general). The period is expressed by the
hyperelliptic integral in $u=\frac{|\zeta|^{2}}{\kappa_{1}^{2}}$:
\[
\frac{T}{T_{u}}=\frac{1}{2\pi}\oint \frac{-9(u+3)\;{\rm
sign}(\xi\eta)}{\sqrt{27c^{3}(u+1)^{2}-2(u+3)^{3}}\sqrt{2(u+3)^{3}-27c^{3}(u-1)^{2}}}\D u.
\]
Novikov gave this expression in the variable
$b=\frac{6}{u+3}=\frac{3\kappa_{1}^{2}|z_{2}-z_{3}|^{2}}{J}$ which always remains
between 0 and 2 but he did not integrate it:
\[
\frac{T}{T_{u}}=\frac{3}{4\pi c^{3}}\oint \frac{{\rm
sign}(\xi\eta)}{\sqrt{f(b)}\sqrt{-g(b)}}\D b
\]
with $f(b)\equiv b(b-3)^{2}-\frac{4}{c^{3}}$, $g(b)\equiv
b(b-\frac{3}{2})^{2}-\frac{1}{c^{3}}$. This hyperelliptic integral happens to be
reducible to an elliptic integral (Bolza, 1898, mentioned in the tables of
Gr\"{o}bner and Hofreiter, 1965) of the variable $z=\frac{g(b)}{3b}$, due to the
relations:
\begin{gather*}
\varphi(z)\equiv
z^{3}-\frac{3}{2}z^{2}+\left(\frac{9}{16}-\frac{3}{2c^{3}}\right)z+\frac{1}{8c^{3}}-
\frac{1}{4c^{6}}=\frac{f(b)[h(b)]^{2}}{27b^{3}},\\[4pt]
\frac{\D z}{\D b}=\frac{6h(b)}{9b^{2}}
\end{gather*}
with $h(b)\equiv b^{3}-\frac{3}{2}b^{2}+\frac{1}{2c^{3}}$; this gives for the
period:
\[
\frac{T}{T_{u}}=\frac{1}{8\pi c^{3}}\oint \frac{{\rm sign}(\xi\eta){\rm
sign}(h(b))}{\sqrt{-z\varphi(z)}}\D z
\]

Let us call $b_{1}<b_{2}<b_{3}$ the zeros of $f$, $b_{4}<b_{5}<b_{6}$ those of $g$
($b_{4}$ and $b_{5}$ are not real for $1<c^{3}<2$), $b_{7}<b_{8}<b_{9}$ those of
$h$ and $z_{1}<z_{2}<z_{3}$ those of $\varphi$. The correspondence is
$(b_{1},b_{9})\to z_{1}$, $(b_{2},b_{8})\to z_{2}$, $(b_{3},b_{7})\to z_{3}$,
which gives the following values of the period for the two domains:
\begin{gather*}
1<c^{3}<2\enskip{:}\enskip {\rm sign}(K)\oint =2\int_{b_{1}}^{b_{2}}\D b=6\int_{z_{1}}^{z_{2}}\D z,\\[3pt]
b_{7}<b_{1}<b_{8}<b_{9}<b_{2}<b_{6}<b_{3}\\[3pt]
\frac{T}{|T_{u}|}=\frac{3}{2\pi c^{3}}\frac{K(k)}{\sqrt{(z_{3}-z_{2})(-z_{1})}},\quad k^{2}=\frac{z_{3}(z_{2}-z_{1})}{(z_{3}-z_{2})(-z_{1})},\quad z_{1}<z_{2}<0<z_{3}\\[3pt]
2<c^{3}\enskip{:}\enskip
b_{7}<b_{1}<b_{4}<b_{8}<b_{5}<b_{9}<b_{6}<b_{2}<b_{3},\quad z_{1}<0<z_{2}<z_{3}.
\end{gather*}

Two equivalent expressions lead to the period, according to whether $\zeta$ turns
around $M_{1}$ or another vortex:
\begin{gather*}
\hspace*{-.05pc}\hbox{($\zeta$ around $M_{1}$)}\hspace*{2.5pc}:\quad {\rm sign}(K)\oint=4\int_{b_{1}}^{b_{4}}\D b=4\int_{z_{1}}^{o}\D z\\[4pt]
\hbox{($\zeta$ around $M_{2}$ or $M_{3}$)}:\quad {\rm sign}(K)\oint=2\int_{b_{5}}^{b_{6}}\D b=4\int_{z_{1}}^{o}\D z\\[4pt]
\frac{T}{|T_{u}|}=\frac{K(k)}{\pi c^{3}\sqrt{z_{3}(z_{2}-z_{1})}},\quad
k^{2}=\frac{-z_{1}(z_{3}-z_{2})}{z_{3}(z_{2}-z_{1})}
\end{gather*}
$\underline{\vec{\kappa}=(-1,1,1,)K}$ where three generic situations exist. We
assume $K>0$.


The $\zeta$ curves are bicircular quartics of genus one and we derive below the
normal forms of the scattering angle and the period of the reduced motion:
\begin{eqnarray*}
\Delta \varphi_{1}&=&\oint\frac{(u+u_{o})\;{\rm sign}(\xi\eta)}{2u\sqrt{(u-u_{-})(u-u_{+})}\sqrt{(u-1)(u_{o}-u)}}\D u\\[4pt]
\frac{T}{T_{u}}&=&\frac{1}{2\pi}\oint\frac{4\;{\rm
sign}(\xi\eta)}{c(u-1)\sqrt{(u-u_{-})(u-u_{+})}\sqrt{(u-1)(u_{o}-u)}}\D u
\end{eqnarray*}
with the notations $u=|\frac{\zeta}{K}|^{2}$, $u_{o}=1+\frac{8}{c}$,
$u_{\pm}=\frac{4}{c}-1\pm \frac{4}{c}$ $\sqrt{1-c}$. The variable $b$ in Aref is
related to $u$ by $b=\frac{6}{1-u}$. $K, E$ and $\Pi$ are the complete elliptic
integrals of the first, second and third kind, the last one being defined
as\footnote{In the tables of Gradshteyn and Ryzhik $4^{{\rm th}}$ edition, the
definition 8.111.4 is not consistent with the rest of the book; many formulae
concerning elliptic integrals are wrong, among them 3.132.5, 3.132.6, 3.138.8,
3.148.1.}
\[
\Pi(n,k)=\int_{0}^{1}\frac{\D x}{(1-nx^{2})\sqrt{(1-x^{2})(1-k^{2}x^{2})}}
\]
First regime (exchange scattering). $-8<c<1$. $\oint {\rm
sign}(\xi\eta)\D u=2\int_{1}^{u_{-}}\D u$.

There is no discontinuity for $c=0$ where $\Delta \varphi_{1}$ evaluates to $\pi$.
\begin{eqnarray*}
&&-8<c<0\enskip{:}\enskip \Delta \varphi_{1}=\frac{-2}{\sqrt{(1-u_{o})(u_{-}-u_{+})}}\\[5pt]
&&\hspace*{5.2pc}\left[\left(1+\frac{u_{0}}{u_{+}}\right)K(k)+\frac{u_{o}(u_{+}-1)}{u_{+}}\Pi(n,k)\right]\\[5pt]
&&\qquad {\rm with}\ k^{2}=\frac{(1-u_{-})(u_{o}-u_{+})}{(1-u_{o})(u_{-}-u_{+})},\quad n=\frac{u_{+}(u_{-}-1)}{u_{-}-u_{+}},\\[5pt]
&&0<c<1\enskip{:}\enskip \Delta \varphi_{1}=\frac{2}{\sqrt{(u_{o}-u_{-})(u_{+}-1)}}\left[2K(k)+\frac{8}{c}\Pi(n,k)\right]\\[5pt]
&&\qquad {\rm with}\ k^{2}=\frac{(u_{o}-u_{+})(u_{-}-1)}{(u_{o}-u_{-})(u_{+}-1)},
\quad n=\frac{u_{o}(1-u_{-})}{u_{o}-u_{-}}
\end{eqnarray*}

\noindent Second regime (direct scattering). $c<-8$ or $1<c$. $\oint {\rm
sign}(\xi\eta)\D u=2\int_{1}^{u_{o}}\D u$
\begin{eqnarray*}
&&c<-8\enskip{:}\enskip \Delta \varphi_{1}=\frac{-2}{\sqrt{(1-u_{-})(u_{o}-u_{+})}}\\[4pt]
&&\hspace*{5.2pc}\left[\left(1+\frac{u_{0}}{u_{+}}\right)K(k)+\frac{(u_{+}-1)u_{o}}{u_{+}}\Pi(n,k)\right]\\[4pt]
&&\qquad {\rm with}\ k^{2}=\frac{(1-u_{o})(u_{-}-u_{+})}{(1-u_{-})(u_{o}-u_{+})},\quad n=\frac{u_{+}(u_{o}-1)}{u_{o}-u_{+}},\\[4pt]
&&1<c\enskip{:}\enskip \Delta \varphi_{1}=(u_{o}^{-\frac{1}{4}}-u_{o}^{\frac{1}{4}})K(k)+\frac{(1+\sqrt{u_{o}})^{2}}{2}\Pi(n,k)\\[4pt]
&&\qquad {\rm with}\
k^{2}=\frac{(u_{o}-1)^{2}-4(\sqrt{u_{o}}-1)^{2}}{16\sqrt{u_{o}}}, \quad
n=-\frac{(1-\sqrt{u_{o}})^{2}}{4\sqrt{u_{o}}}
\end{eqnarray*}
Third regime (biperiodic). $0<c<1$ $\oint {\rm
sign}(\xi\eta)\D u=4\int_{u_{+}}^{u_{o}}\D u$
\begin{eqnarray*}
&&\frac{T}{T_{u}}=\frac{2}{\pi}\sqrt{\frac{2(u_{o}-1)}{(u_{-}-1)(u_{o}-u_{-})}}\left[K(k)-\frac{u_{o}-u_{-}}{6(u_{o}-1)}E(k)\right]\\
&&\Delta \varphi_{1}=\frac{4}{\sqrt{(u_{-}-1)(u_{o}-u_{-})}}\left[2K(k)-\frac{8}{c+8}\Pi(n,k)\right]\\
&&\qquad {\rm with}\ k^{2}=\frac{(u_{-}-1)(u_{o}-u_{+})}
{(u_{+}-1)(u_{o}-u_{-})},\quad n=\frac{u_{+}-u_{o}}{u_{o}(u_{+}-1)}
\end{eqnarray*}
$\underline{\vec{\kappa}=(-1,2,2)\frac{K}{3}}$, a case with two possible regimes
(biperiodic, expanding).

The $\zeta$ trajectories are the Cassini ovals, whose genus is one. The period for
instance is given by
\[
T=\left(\frac{4\pi^{2}J}{K^{3}}\right)\frac{\alpha}{2\pi}\oint\frac{-27\;{\rm
sign}(\xi\eta)}{4(u-3)^{2}\sqrt{(u+1)^{2}-\alpha}\sqrt{\alpha-(u-1)^{2}}}\D u
\]
with $u=|\frac{\zeta}{\kappa_{2}}|^{2}$,
$\alpha=|(\frac{\zeta}{\kappa_{2}})^{2}-1|^{2}=16$ $e^{18\pi E/K^{2}}$. Its
reduced form is not very compact and we shall not give it here.

\section*{Appendix II}

\section*{Stability of the relative equilibria} 

In order to obtain the shape of the $\zeta$ trajectories in the vicinity of the relative
equilibria we must determine whether they are of elliptic or hyperbolic nature.

The points $\infty,-\kappa_{2},\kappa_{3}$ are elliptic, neighbouring orbits are
circles described with a uniform circular motion of period:
\[
T=\frac{4\pi^{2}J}{(K-\kappa_{j})^{2}\kappa_{j}}\left(\frac{Qc}{\kappa_{j}(K-\kappa_{j})}\right)^{-\frac{Q\kappa_{j}}{\kappa_{1}\kappa_{2}\kappa_{3}}}
\]

We now assume that $\zeta_{o}$ is the affix of an ordinary relative equilibrium
(the case of two coincident r.e. is studied elsewhere in the paper), which implies
$J \ne 0$ and we study the vicinity of the equilateral triangles and of the
aligned configurations.

By writing $f(\zeta,\bar{\zeta})$ for the right-hand side of Equation (10), the
small motions of a $\zeta$ point in the vicinity of a stationary point $\zeta_{o}$
are ruled by:
\[
2\pi i\frac{\D \bar{\zeta}}{\D t}=(\zeta-\zeta_{o})\frac{\partial f}{\partial \zeta}
(\zeta_{o},\bar{\zeta}_{o})+(\overline{\zeta-\zeta_{o}})\frac{\partial f}{\partial
\bar{\zeta}}(\zeta_{o},\bar{\zeta}_{o}),
\]
or, in real matricial notation:
\[
2\pi\frac{\D}{\D t}\binom{\xi}{\eta}=M\binom{\xi-\xi_{o}}{\eta-\eta_{o}}=
\left(\begin{array}{@{}l@{\quad}l@{}}
\alpha'+\beta'&\alpha-\beta\\
\alpha+\beta&-\alpha'+\beta'\\
\end{array}\right)\binom{\xi-\xi_{o}}{\eta-\eta_{o}}
\]
with $\alpha+i\alpha'=\frac{\partial f}{\partial \zeta}(\zeta_{o},\bar{\zeta}_{o}) =\mu$,
$\beta+i\beta'=\frac{\partial f}{\partial \bar{\zeta}}(\zeta_{o},\bar{\zeta}_{o})=\nu$.

The stability condition is: ${\rm tr}(M)=0$, ${\rm det}(M)>0$. We find:
\[
\frac{-(\kappa_{1}|\zeta|^{2}+\kappa_{2}\kappa_{3}K)}{J(\zeta+\kappa_{2})(\zeta-\kappa_{3})}=
\frac{\mu}{2|\zeta|^{2}+d\bar{\zeta}-sK}=\frac{\nu}{\zeta^{2}+d\zeta-Q}
\]
with the condition (17). The trace of $M$ is therefore zero. If $\zeta_{o}$ is
elliptic, then the small motions have the period $4\pi^{2}/\sqrt{{\rm det}(M)}$.
We now divide the study according to the two types of stationary points.

\medskip\noindent(a) \underline{The equilateral triangles}

${\rm det}(M)=\frac{3Q^{3}}{J^{2}}$. The stability condition is $Q>0$ and, when
this is fulfilled, the period of the small motions is
$\frac{4\pi^{2}J}{\sqrt{3}Q^{3/2}}$; comparing with the period of the absolute
motion which will be derived later, we find:
\[
\left(\frac{T_{r}}{T_{a}}\right)^{2}=1-\frac{1}{2K^{2}}\sum_{j}\sum_{\ell >
j}(\kappa_{j}-\kappa_{\ell})^{2}.
\]
The overall rotation is therefore quicker than the small motions with equality
only for identical strengths.

\medskip\noindent(b) \underline{The aligned configurations}
\[
{\rm det}\
M=|\nu|^{2}-|\mu|^{2}=\left(\frac{\kappa_{1}\zeta^{2}+\kappa_{2}\kappa_{3}K}{J(\zeta^{2}+d\zeta-\kappa_{2}\kappa_{3})}\right)^{2}
F_{1}(\zeta)F_{2}(\zeta)
\]
with the notation
\begin{eqnarray*}
F_{1}(\zeta)&=&-\zeta^{2}+Ks-Q\\
F_{2}(\zeta)&=&3\zeta^{2}+2d\zeta-sK-Q=P'(\zeta)
\end{eqnarray*}
and the condition: $\zeta$ is a real zero of $P$.

The determinant of $M$ changes sign when the resultant of $P$ and $F_{1}F_{2}$
vanishes. We find:
\[
{\rm res}(P,F_{1})=-3s^{2}Q^{2},\quad {\rm res}(P,F_{2})=s^{2}\Delta.
\]
Then we obtain the nature of the aligned configurations in the parameter space:
\begin{eqnarray*}
\hbox{$\Delta > 0$ $(Q < 0)$}&:& \hbox{one stable configuration}\\
\hbox{$\Delta < 0$ and $Q<0$}&:& \hbox{two stable configurations, one unstable}\\
\hbox{$\Delta < 0$ and $Q>0$}&:& \hbox{three unstable configurations}.
\end{eqnarray*}

It is worth observing that in the present problem the nonlinear stability is the same as
the linear one.


\begin{thebibliography}{00}
\bibitem{Aref} 
H.~Aref (1979), Motion of three vortices, Phys.~Fluids {\it 22}, 393--400.

\bibitem{Bateman}
Bateman manuscript project, (1953), Higher transcendental functions, vol.~II,
chapter XIII; A.~Erd\'{e}lyi Editor, Mc Graw Hill. 

\bibitem{ConteThese}
R.~Conte (1979), Th\`{e}se d'Etat, Universit\'{e} de Paris VI. 

\bibitem{AMP}
J.M.~Drouffe, AMP language, (1976), same address as authors.

\bibitem{ET}
S.F.~Edwards and J.B.~Taylor (1974), Negative temperature states of
two-dimensional plasmas and vortex fluids, Proc. Roy. Soc. London, {\it A 336},
\hbox{257--271}.

\bibitem{GH}
W.~Gr\"{o}bner and Hofreiter (1965), Integraltafel, vol.~4, Springer-Verlag.

\bibitem{LLMachanics}
L.~Landau and E.~Lifchitz (1960), Mechanics, Pergamon Press.

\bibitem{LP}
T.S.~Lundgren and Y.B.~Pointin (1977), Statistical mechanics of two-dimensional
vortices, Journal of statistical physics, {\it 17}, 323--325.

\bibitem{Mayer}
A.M.~Mayer (1878), Floating magnets, Nature, {\it 18}, 258.

\bibitem{Novikov}
E.A.~Novikov (1975), Dynamics and statistics of a system of vortices, JETP {\it
41},~937--943.

\bibitem{Onsager}
L.~Onsager (1949), Statistical hydrodynamics, Nuovo Cimento, {\it 6} suppl.,
279--287.

\bibitem{HP}
H.~Poincar\'{e} (1893), Th\'{e}orie des tourbillons, pages 77--84, Deslis
fr\`{e}res, Paris.

\bibitem{Seyler}
C.E.~Seyler, Jr.~(1974), Partition function for a two-dimensional plasma in the
random phase approximation, Phys.~Rev.~Letters {\it 32}, 515--517.

\bibitem{Smale}
S.~Smale (1970), Topology and mechanics, Inventiones math., {\it 10}, 305--331 and
{\it 11}, 45--64.

\bibitem{Thomson}
W.~Thomson (1878), Floating magnets (illustrating vortex-systems), Nature, {\it
18},~13--14.

\end{thebibliography}
\end{document}